\documentclass[superscriptaddress,aps,prl,showkeys,showpacs,nofootinbib,twocolumn,10pt]{revtex4}
\usepackage[OT1,T1]{fontenc}
\usepackage{graphicx}
\usepackage{rotating}
\usepackage[usenames]{color}
\usepackage{epsfig}
\usepackage{amssymb,amsmath}
\usepackage{epstopdf}
\usepackage{subfig}
\usepackage{lineno,hyperref}
\usepackage{enumitem}

\definecolor{grey}{rgb}{0.7,0.7,0.7}
\definecolor{black}{rgb}{0,0,0}
\def \grey{\color{grey} }

\newcommand{\di}{i}

\hypersetup{pdfpagemode=FullScreene,pdfstartview={XYZ null null 1.10}}

\begin{document}

\title{Differential Cross Sections for Neutron-Proton Scattering in the Region
of the $d^*(2380)$ Dibaryon Resonance}
\date{\today}

\newcommand*{\IKPUU}{Division of Nuclear Physics, Department of Physics and 
 Astronomy, Uppsala University, Box 516, 75120 Uppsala, Sweden}
\newcommand*{\York}{Department of Physics, University of York, Heslington,
  York, YO10 5DD, UK}
\newcommand*{\ASWarsN}{Department of Nuclear Physics, National Centre for 
 Nuclear Research, ul.\ Hoza~69, 00-681, Warsaw, Poland}
\newcommand*{\IPJ}{Institute of Physics, Jagiellonian University, ul.\ 
 Reymonta~4, 30-059 Krak\'{o}w, Poland}
\newcommand*{\PITue}{Physikalisches Institut, Eberhard--Karls--Universit\"at 
 T\"ubingen, Auf der Morgenstelle~14, 72076 T\"ubingen, Germany}
\newcommand*{\Kepler}{Kepler Center for Astro and Particle Physics, University
 of T\"ubingen, Auf der Morgenstelle~14, 72076 T\"ubingen, Germany}
\newcommand*{\MS}{Institut f\"ur Kernphysik, Westf\"alische 
 Wilhelms--Universit\"at M\"unster, Wilhelm--Klemm--Str.~9, 48149 M\"unster, 
 Germany}
\newcommand*{\ASWarsH}{High Energy Physics Department, National Centre for 
 Nuclear Research, ul.\ Hoza~69, 00-681, Warsaw, Poland}
\newcommand*{\ASWarsHE}{High Energy Physics Division, National Centre for 
 Nuclear Research, 05-400 Otwock-wierk, Poland}
\newcommand*{\IITB}{Department of Physics, Indian Institute of Technology 
 Bombay, Powai, Mumbai--400076, Maharashtra, India}
\newcommand*{\PGI}{Peter Gr\"unberg Institut, Forschungszentrum J\"ulich,
  52425 J\"ulich, Germany}
\newcommand*{\ILP}{Institut f\"ur Laser- und Plasmaphysik, Heinrich-Heine
  Universit\"at D\"usseldorf, 40225 D\"usseldorf, Germany}
\newcommand*{\IKPJ}{Institut f\"ur Kernphysik, Forschungszentrum J\"ulich, 
 52425 J\"ulich, Germany}
\newcommand*{\Bochum}{Institut f\"ur Experimentalphysik I, Ruhr--Universit\"at 
 Bochum, Universit\"atsstr.~150, 44780 Bochum, Germany}
\newcommand*{\ZELJ}{Zentralinstitut f\"ur Engineering, Elektronik und 
 Analytik, Forschungszentrum J\"ulich, 52425 J\"ulich, Germany}
\newcommand*{\Erl}{Physikalisches Institut, 
 Friedrich--Alexander--Universit\"at Erlangen--N\"urnberg, 
 Erwin--Rommel-Str.~1, 91058 Erlangen, Germany}
\newcommand*{\ITEP}{Institute for Theoretical and Experimental Physics, State 
 Scientific Center of the Russian Federation, Bolshaya Cheremushkinskaya~25, 
 117218 Moscow, Russia}
\newcommand*{\Giess}{II.\ Physikalisches Institut, 
 Justus--Liebig--Universit\"at Gie{\ss}en, Heinrich--Buff--Ring~16, 35392 
 Giessen, Germany}
\newcommand*{\IITI}{Department of Physics, Indian Institute of Technology 
 Indore, Khandwa Road, Indore--452017, Madhya Pradesh, India}
\newcommand*{\Aachen}{III.~Physikalisches Institut~B, Physikzentrum, 
 RWTH Aachen, 52056 Aachen, Germany}
\newcommand*{\HepGat}{High Energy Physics Division, Petersburg Nuclear Physics 
 Institute, Orlova Rosha~2, Gatchina, Leningrad district 188300, Russia}
\newcommand*{\HeJINR}{Veksler and Baldin Laboratory of High Energiy Physics, 
 Joint Institute for Nuclear Physics, Joliot--Curie~6, 141980 Dubna, Russia}
\newcommand*{\Katow}{August Che{\l}kowski Institute of Physics, University of 
 Silesia, Uniwersytecka~4, 40-007, Katowice, Poland}
\newcommand*{\IFJ}{The Henryk Niewodnicza{\'n}ski Institute of Nuclear 
 Physics, Polish Academy of Sciences, 152~Radzikowskiego St, 31-342 
 Krak\'{o}w, Poland}
\newcommand*{\NuJINR}{Dzhelepov Laboratory of Nuclear Problems, Joint 
 Institute for Nuclear Physics, Joliot--Curie~6, 141980 Dubna, Russia}
\newcommand*{\KEK}{High Energy Accelerator Research Organisation KEK, Tsukuba, 
 Ibaraki 305--0801, Japan} 
\newcommand*{\ASLodz}{Astrophysics Division, National Centre for Nuclear
  Research, Box 447, 90-950 {\L}\'{o}d\'{z}, Poland
}
\newcommand*{\Tomsk}{Department of Physics, Tomsk State University, 36~Lenin 
 Ave., Tomsk, 634050 Russia}

\author{P.~Adlarson}    \affiliation{\IKPUU}
\author{W.~Augustyniak} \affiliation{\ASWarsN}
\author{W.~Bardan}      \affiliation{\IPJ}
\author{M.~Bashkanov}   \affiliation{\York}
\author{F.S.~Bergmann}  \affiliation{\MS}
\author{M.~Ber{\l}owski}\affiliation{\ASWarsH}
\author{H.~Bhatt}       \affiliation{\IITB}
\author{M.~B\"uscher}\affiliation
{\PGI}\affiliation{\ILP}
\author{H.~Cal\'{e}n}   \affiliation{\IKPUU}
\author{I.~Ciepa{\l}}   \affiliation{\IPJ}
% ...

\newcommand*{\Delhi}{Department of Physics and Astrophysics, University of 
 Delhi, Delhi--110007, India}
\newcommand*{\Mainz}{Institut f\"ur Kernphysik, Johannes 
 Gutenberg--Universit\"at Mainz, Johann--Joachim--Becher Weg~45, 55128 Mainz, 
 Germany}
\newcommand*{\Bern}{Albert Einstein Center for Fundamental Physics, University 
 of Bern, Sidlerstrasse~5, 3012 Bern, Switzerland}
\newcommand*{\Sheff}{Department of Physics and Astronomy, University of 
 Sheffield, Hounsfield Road, Sheffield, S3 7RH, United Kingdom}
\newcommand*{\Roma}{Dipartimento di Fisica dell Universita Sapienza, Roma, Italy
and INFN Sezione di Roma, Roma, Italy}

\collaboration{WASA-at-COSY Collaboration}\noaffiliation

\newcommand*{\GW}{Data Analysis Center at the Institute for Nuclear Studies,
  Department of Physics, The George Washington University, Washington,
  D.C. 20052, U.S.A.}
\author{R. L. Workman}     \affiliation{\GW}
\author{W. J. Briscoe}     \affiliation{\GW}
\author{I. I. Strakovsky}  \affiliation{\GW}

\collaboration{SAID Data Analysis Center}\noaffiliation

\author{A.~\v{S}varc} \affiliation{Rudjer Bo\v{s}kovi\'{c} Institute, Bijeni\v{c}ka cesta 54, P.O.
Box 180, 10002 Zagreb, Croatia}

\affiliation{Tesla Biotech, Mandlova 7, 10002 Zagreb, Croatia}

\begin{abstract}
Differential cross sections have been extracted from exclusive and
kinematically complete high-statistics measurements of quasifree polarized
$\vec{n}p$ scattering performed in the energy region of the $d^*(2380)$
dibaryon resonance covering the the range of beam energies $T_n$ = 0.98 -
1.29 GeV ($\sqrt s$ = 2.32 - 2.44 GeV). The experiment was carried out with
the WASA-at-COSY setup having a polarized deuteron beam impinged on the
hydrogen pellet target and utilizing the quasifree process $dp \to np +
p_{spectator}$. That way the $np$ differential cross section $\sigma(\Theta)$
was measured over a large angular range. The obtained  angular
distributions complement the corresponding analyzing power $A_y(\Theta)$
measurements published previously.
A SAID partial-wave analysis incorporating the new data strengthens the
finding of a resonance pole in the coupled $^3D_3 - ^3G_3$ waves.

\end{abstract}

\pacs{13.75.Cs, 13.85.Dz, 14.20.Pt}

\maketitle

\section{Introduction}

Recently a resonance pole with $I(J^P) = 0(3^+)$ at ($2380\pm10 - i 40\pm5$)
MeV --- denoted $d^*(2380)$ --- was discovered in the coupled $^3D_3 - ^3G_3$  
partial waves of nucleon-nucleon ($NN$) scattering by the SAID partial-wave
analysis based on the full SAID data base and recent analyzing power data
provided by WASA-at-COSY for the lab 
energy range $T_n$ = 1.095 - 1.270 GeV ($\sqrt s$ = 2.36 - 2.43 GeV)
\cite{np,npfull}. The values for this pole coincide with a pronounced
narrow resonance structure previously observed in the total cross
section of the basic isoscalar double-pionic fusion reaction $pn \to d
\pi^0\pi^0$ \cite{mb,MB} at a mass $M \approx$ 2370~MeV with a width of 
$\Gamma \approx$ 70 MeV. From the angular distributions spin-parity $J^P =
3^+$ was deduced \cite{MB}. Additional evidence for $d^*(2380)$
has been found recently in the $pn \to d\pi^+\pi^-$ \cite{isofus}, $pn
\to pp\pi^0\pi^-$ \cite{TS}, $pn \to pn\pi^0\pi^0$ \cite{np00} and $pn \to pn
\pi^+\pi^-$ \cite{hades,stori} reactions. In measurements of the isoscalar
single-pion production cross section no significant decay of this resonance
into the isoscalar $[NN\pi]_{I=0}$ channel has been observed --- providing a
small upper limit \cite{NNpi}. That way all branchings of this resonance
into the hadronic decay channels $NN$, $NN\pi$ and $NN\pi\pi$ have been
extracted \cite{NNpi,BR}. They agree with the decay of a deeply bound
$\Delta\Delta$ system \cite{BR,hcl}, possibly accompanied with a small
admixture of a $D_{12}\pi$ configuration \cite{GalPLB}, where $D_{12}$ denotes
the $I(J^P) = 1(2^+)$ resonance structure near the $\Delta N$ threshold. For a
discussion of the latter see, {\it e.g.}, Ref. \cite{hcl}. Recently also
suggestive evidence for a photo-exitation of $d^*(2380)$ has been found in
measurements of the $\gamma d \to d\pi^0\pi^0$ reaction \cite{ELPH,Basel}.

In addition to the many evidences for the dibaryon resonance $d^*(2380)$ the
estasblishment of its resonance pole in $np$ scattering certainly is of particular
importance. This finding is solely based on the analyzing power data provided
by WASA-at-COSY. Hence it appears highly desireable to supplement this data base
in the region of the $d^*(2380)$ resonance by comprehensive differential cross
section data, since previous measurements mainly covered just either the very
forward \cite{TE} or the backward angle \cite{BI} region.

\section{Experiment}

For the extraction of the differential cross sections in the region of the
$d^*(2380)$ resonance we use the same data set as exploited before for the
extraction of the analyzing powers \cite{np,npfull}. For this purpose the $np$
elastic scattering was measured in the quasifree mode with the WASA detector
including a hydrogen pellet target \cite{CB,wasa} at 
COSY (Forschungsztentrum J\"ulich, Germany) and by using a polarized deuteron beam with an energy of
$T_d$~=~2.27~GeV. That way the full energy range of the conjectured resonance
was covered. Note that we observe here the quasi-free scattering process
$d p
\to np + p_{spectator}$ in inverse kinematics, which allows to detect also the
fast spectator proton in the forward detector of WASA.

Since we deal here with events originating from channels with large cross
section, the trigger was set to at least one hit in the first layer of the
forward range hodoscope. 
For the case of quasifree $np$ scattering this defines two
event classes with each of them having the spectator proton detected in the
forward detector:
\begin{itemize}
\item  scattered proton and scattered neutron both detected in the central
  detector covering the neutron angle region $40^\circ < \Theta_n^{cm} <
  125^\circ$ 
\item scattered proton detected in the forward detector with the scattered
  neutron being unmeasured covering thus $132^\circ < \Theta_n^{cm} < 145^\circ$
\end {itemize}

 That way a large range of neutron scattering angles could be
 covered. 

For each selected event one neutral hit in the central detector was
required. The $pn$ elastic events have been identified by using the kinematic
constraints for opening angle and planarity.

Since by use of the inverse kinematics the spectator proton is in the beam
particle, the deuteron, the spectator is very fast. This allows its detection in
the forward detector and by reconstruction of its kinetic energy and its
direction  
the full four-momentum of
the spectator proton has been determined.

Similarly the four-momentum of the actively scattered proton has been obtained
from its track information in either forward or central detector (in the
latter case the energy information was not retrieved). 

Since the neutron has been detected by a hit in the calorimeter 
(composed of 1012 CsI(Na) crystals) of the central
detector-- associated with no hit in the preceding plastic
scintillator barrel --, only its directional information has been obtained. In
the subsequent kinematical fit the full event could be reconstructed with two
overconstraints in case of the first event class and with three
overconstraints in case of the second event class.



As noted above, we utilize here data, which have been obtained by use of a
polarized beam for the determination of analyzing powers. Hence,
in order not to distort the beam polarization, the magnetic field
of the solenoid in the central detector was switched off in that beamtime.

Whereas in analyzing power measurements detector efficiencies cancel out, the
determination of differential cross sections heavily depends on a precise
knowledge of detector efficiencies. The latter have been determined by
comprehensive Monte-Carlo (MC) simulations of the WASA detector performance
and their cross check against calibration data.

\begin{figure} 
\centering
\includegraphics[width=0.79\columnwidth]{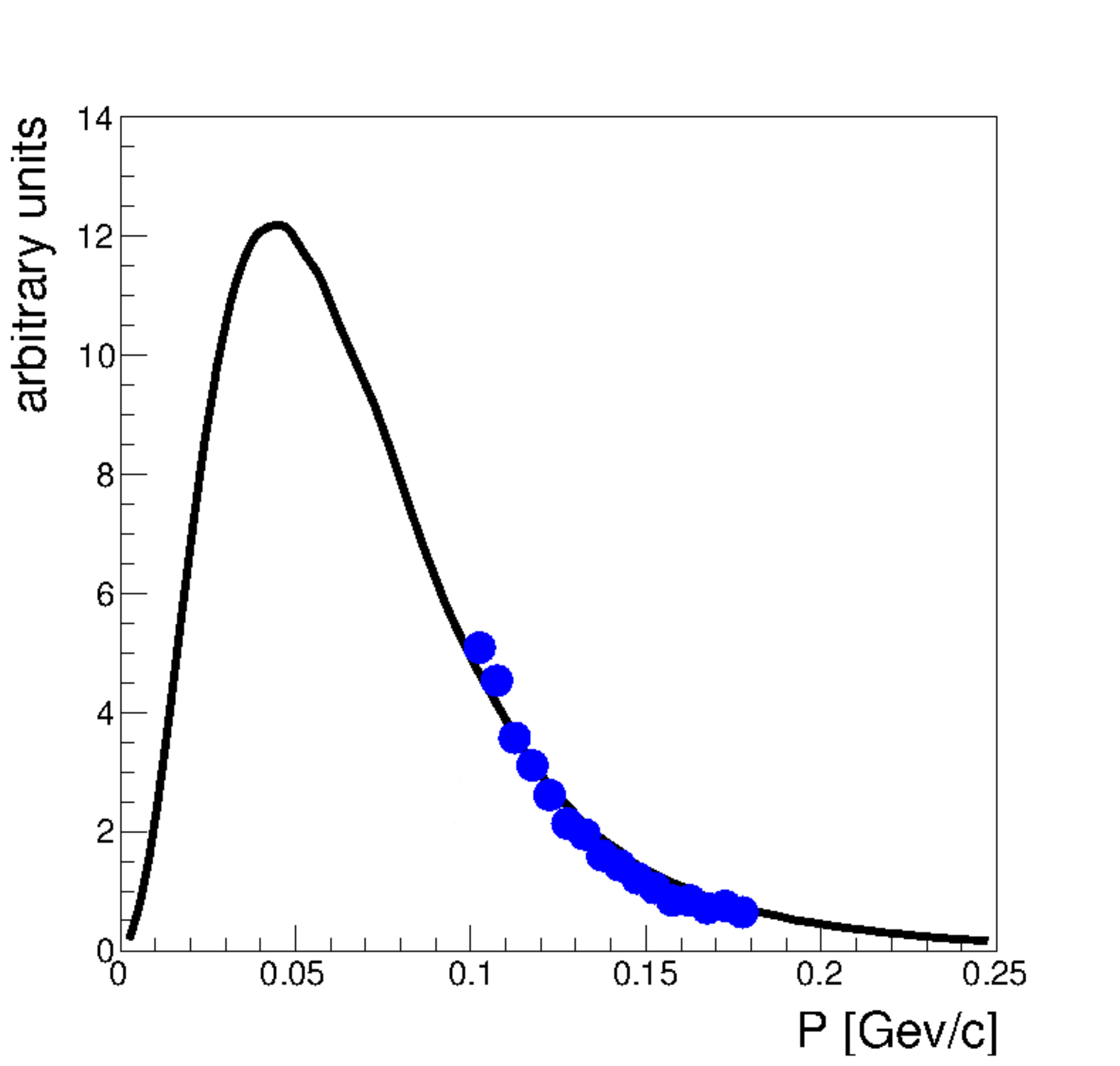}
\caption{\small 
  Distribution of the spectator proton momenta P (in the deuteron rest frame) in
  the $dp \to pn +  
  p_{spectator}$ reaction. Data are
  given by the full circles. The solid line shows 
  the expected distribution for the quasifree process based on the CD Bonn 
  potential \cite{Machleidt} deuteron wavefunction. 
  For the data analysis only events with spectator momenta $P <$ 0.18 GeV/c
  have been used.
}
\label{fig1}
\end{figure}

The momentum distribution of the observed spectator proton in the elastic $np$
scattering process is plotted in Fig.~1 in the deuteron rest frame and
compared with Monte Carlo simulations of the proton momentum distribution 
in the deuteron. 
In these simulations the deuteron wavefunction of CD Bonn potential
\cite{Machleidt} has been used. Because of the beam-pipe
ejectiles can only be detected in the forward detector for lab angles larger
than 2.5 degrees. In order to assure a quasi-free process we omit events
with  spectator momenta larger than 0.18 GeV/c (in the deuteron rest system)
from the subsequent analysis -- similar to what was done in previous work
\cite{MB,isofus,TS}.   

The absolute normalization of our data has been obtained by normalizing our
data at $T_p$ = 1.109 GeV to the backangle data of Bizard {\it et al.}
\cite{BI}. 

For a cross check of the absolut normalization of the $np$ scattering data we
have analyzed the $dp \to np\pi^0 + p_{spectator}$ reaction, which has been
taken in parallel and with the same trigger.
Since there are no high-quality data for the
$np\pi^0$ channel at the energy of interest here, we have used the following
isospin relation for the total cross sections \cite{Bystricky}

\begin{eqnarray}
2 \sigma(np \to np\pi^0) =&&\sigma(pp \to pn\pi^+) + \\
&&2\sigma(np \to pp\pi^-) - 2 \sigma (pp \to pp\pi^0)  \nonumber
\end{eqnarray}

Using the values $17 \pm 2.2$ mb \cite{hades}, $2.5 \pm 0.2$ mb and $4.0 \pm
0.3$ mb \cite{NNpi} for the
total cross sections of the $pp \to pn\pi^+$, $np \to pp\pi^-$ and $pp \to
pp\pi^0$ reactions, respectively, we arrive at a total cross section $6 \pm 1$
mb for the $np \to np\pi^0$ reaction. Using the absolute normalization as
obtained from the adjustment of our $np$ data to those of Bizard {\it et al.},
we arrive at 7 mb for the $np \to np\pi^0$ reaction --- in good agreement with
the value obtained from the isospin relation.

\section{Experimental Results}

Due to the Fermi motion of the nucleons bound in the beam deuteron, the
measurement of the quasi-free $np$ scattering process covers a range of
energies in the $np$ system. Meaningful statistics could be collected for
the range of $np$ center-of-mass energies 
2.32 $< \sqrt s <$ 2.44 
corresponding to $T_n$ = 0.98 - 1.29 GeV. 

By taking the measured spectator four-momentum into
account and reconstructing that way the effective $\sqrt s$ for each event, we
obtain angular distributions for six $\sqrt s$ 
bins as shown in Fig.~2. Our data agree well with previous experimental
results from Saturne for backward \cite{BI} angles in the overlap
region. Where overlapping our results are also in reasonable agreement
with the old Birmingham data \cite{MU},  which were discarded in previous SAID
analyses, though they were taken over nearly the full angular range at $T_p$ =
0.991 GeV . Our data are also in good agreement with
the old Berkeley data taken at $T_p$ = 1.243 GeV over the full forward angular
range \cite{PE}. 

\begin{figure*}
\centering
\includegraphics[width=0.99\columnwidth]{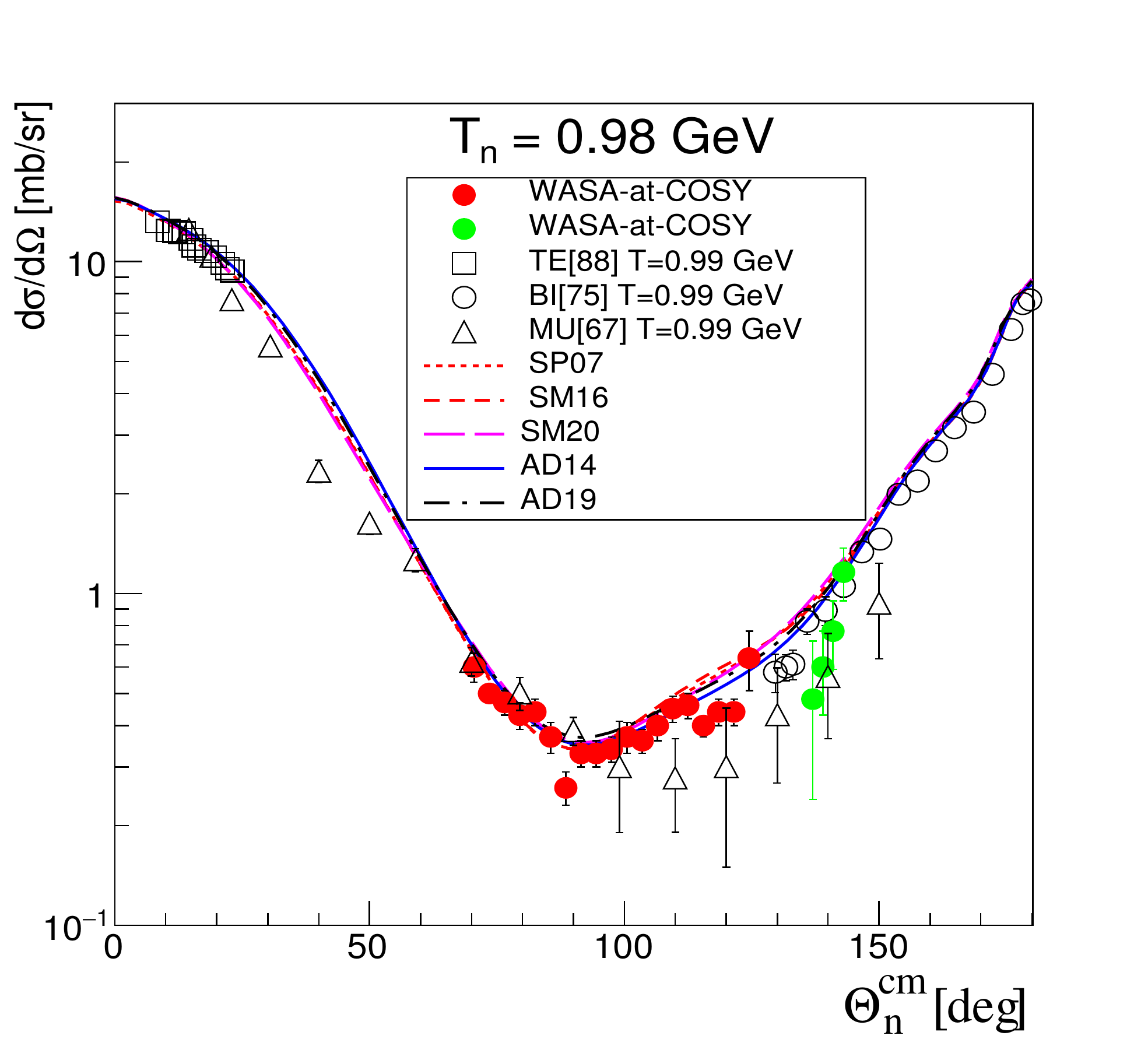}
\includegraphics[width=0.99\columnwidth]{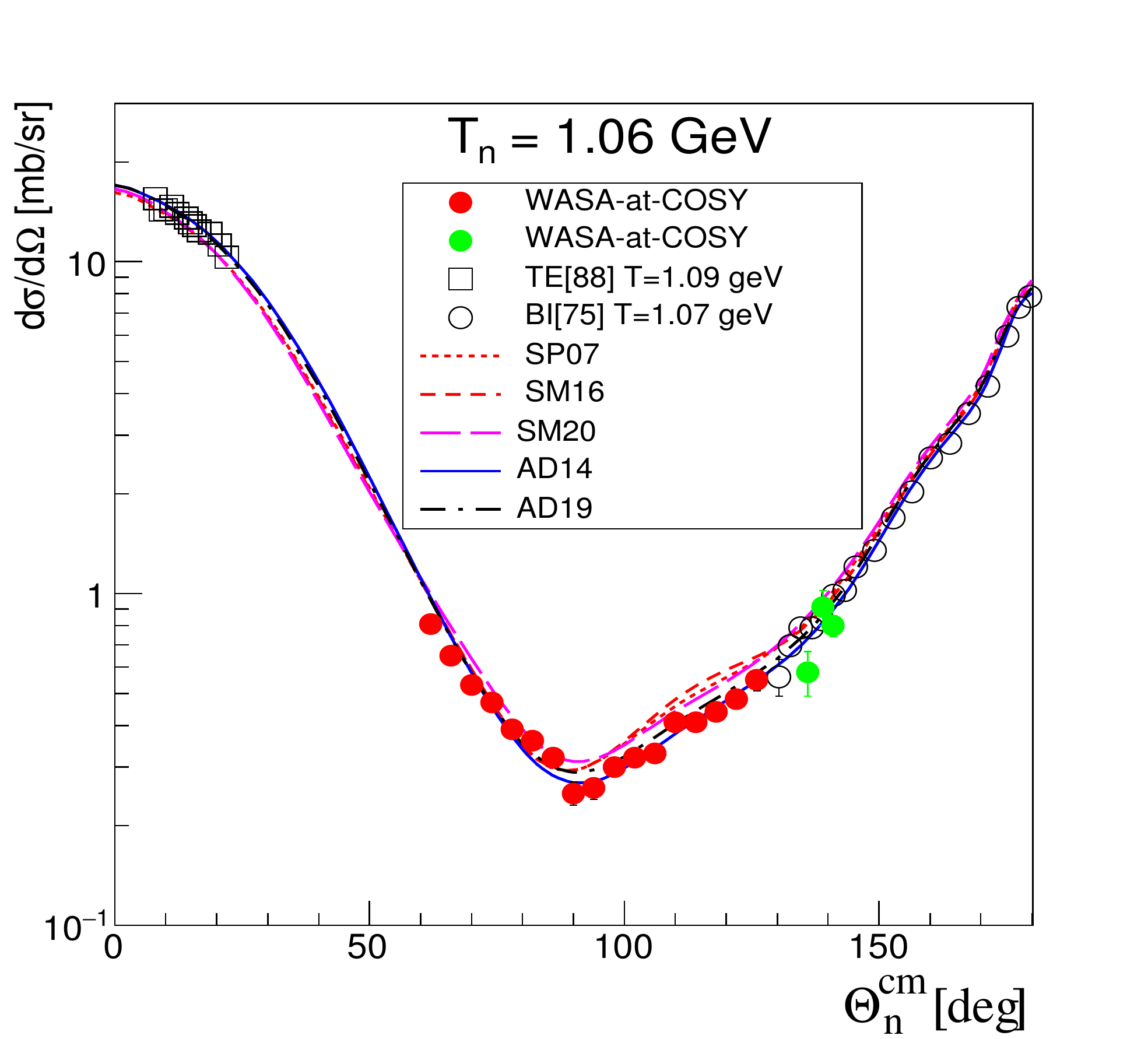}
\includegraphics[width=0.99\columnwidth]{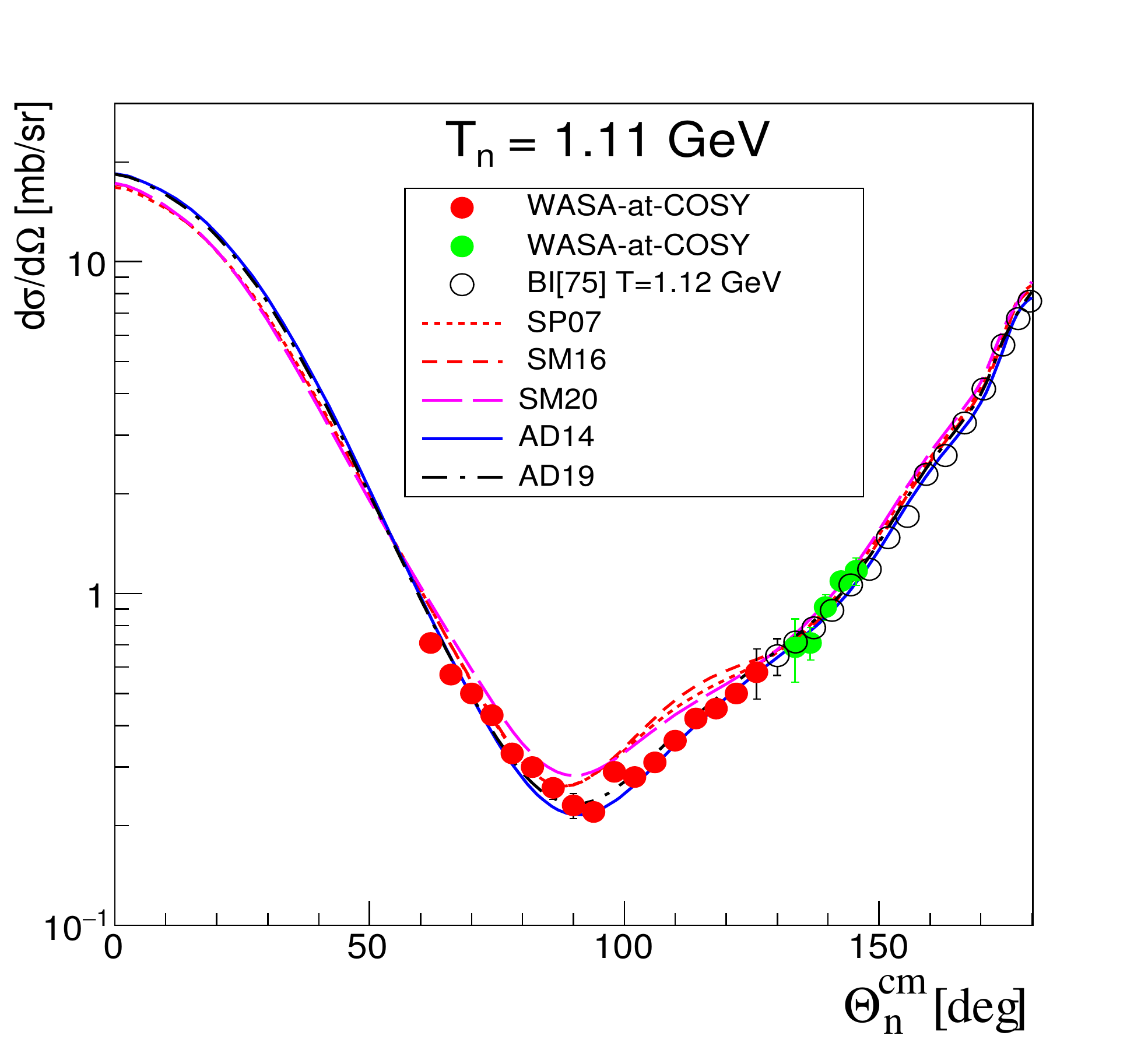}
\includegraphics[width=0.99\columnwidth]{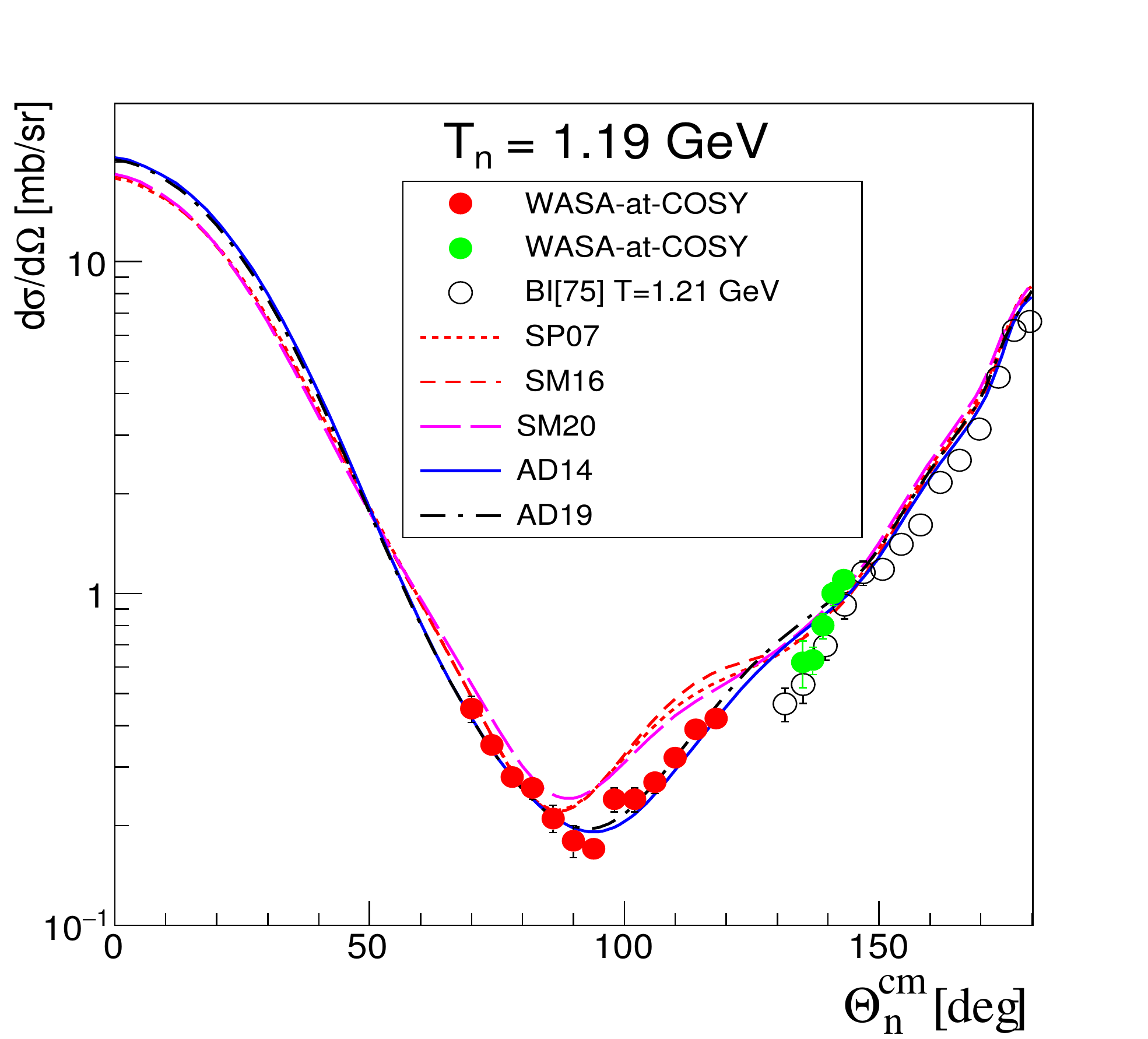}
\includegraphics[width=0.99\columnwidth]{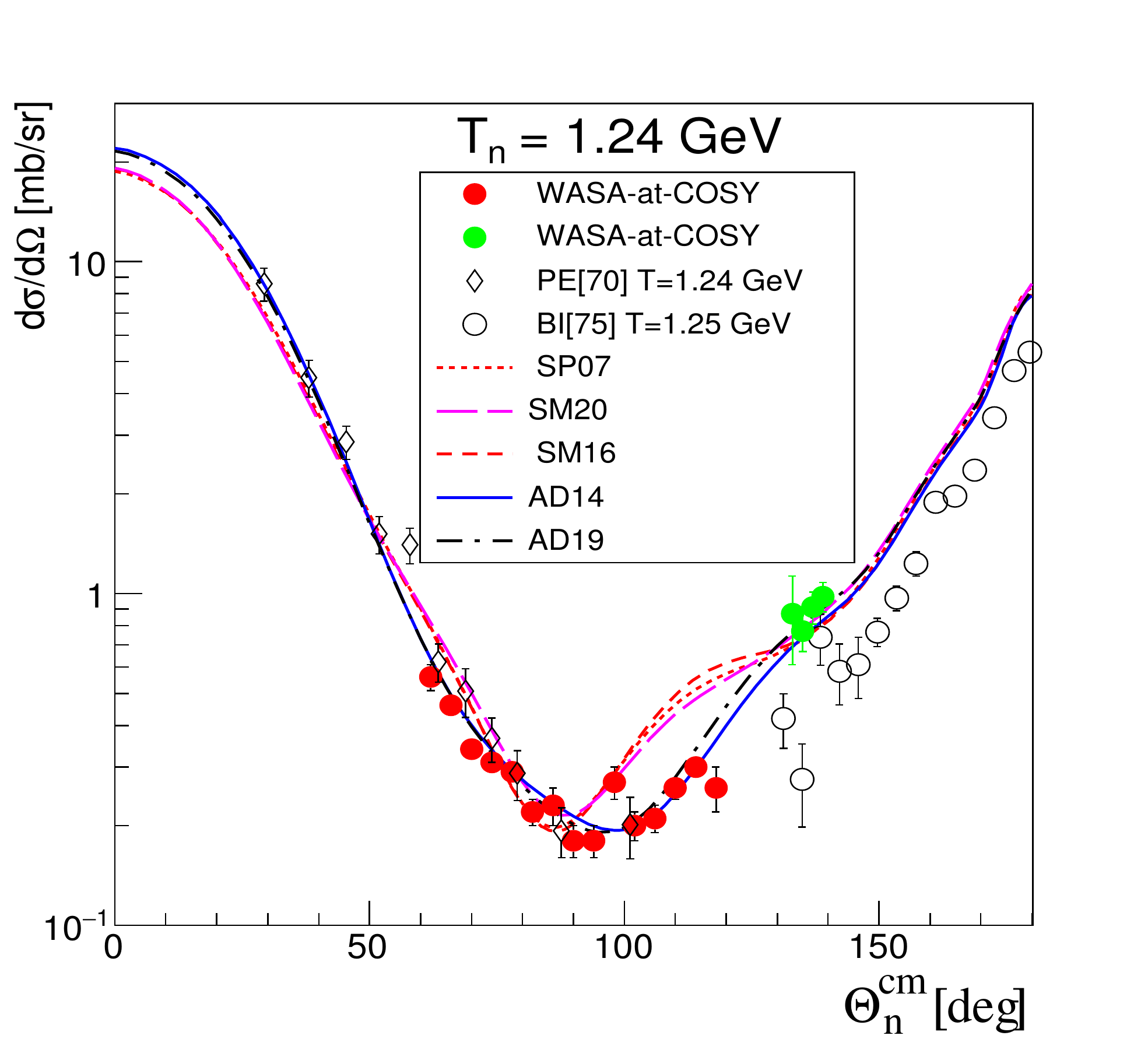}
\includegraphics[width=0.99\columnwidth]{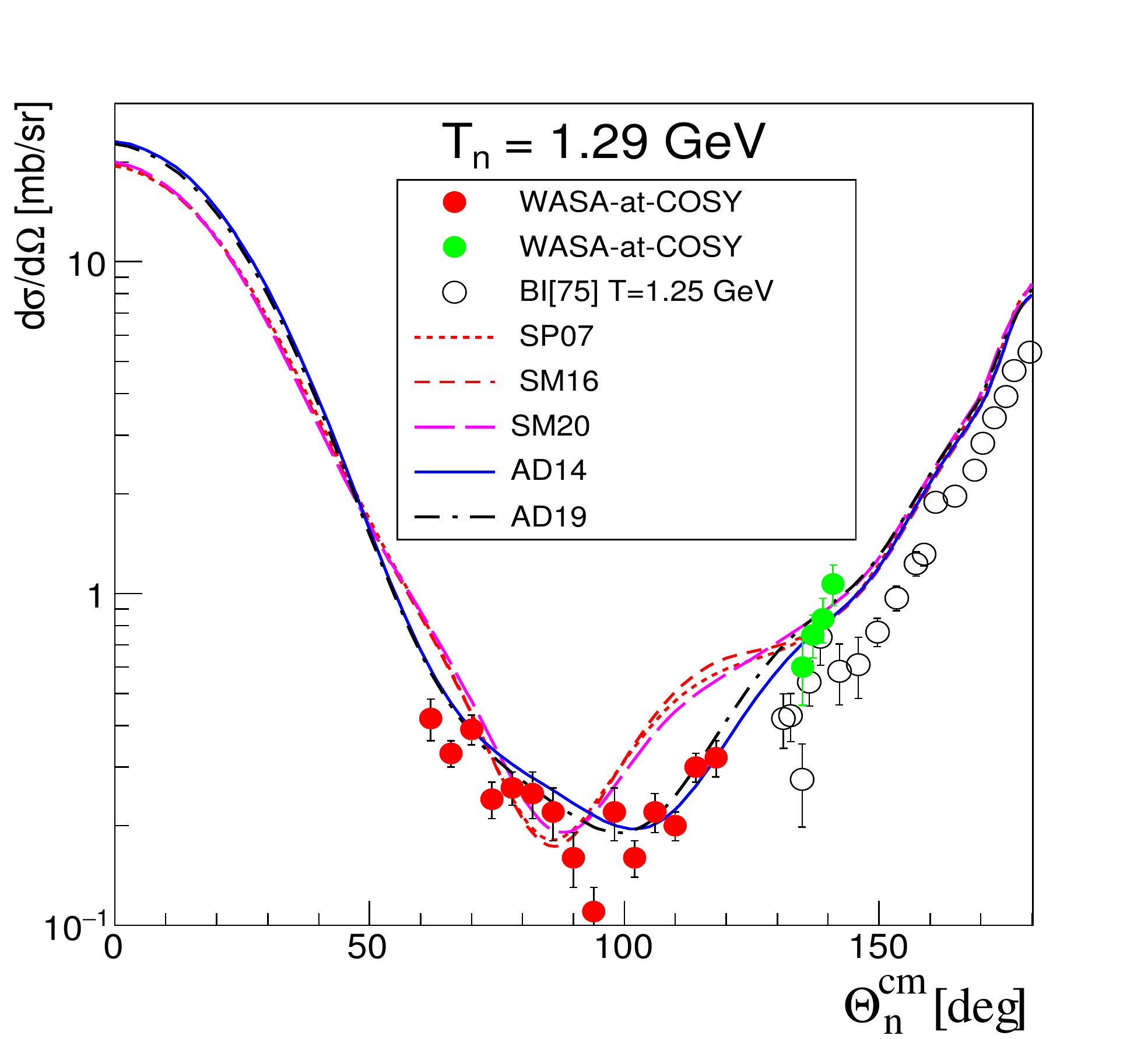}
\caption{\small (Color online) Differential cross sections for elastic $np$
  scattering at $T_n$ = 0.98, 1.06, 1.11, 1.19, 1.24 and 1.29 GeV
  corresponding to $\sqrt s$ = 2.32, 2.35, 2.37, 2.40, 2.42 and 2.44
  GeV. The full symbols denote results from this work taking into account the
  spectator four-momentum information. Open symbols refer to previous
  measurements: for "TE[88]" see Ref.\cite{TE}, for "BI[75]" see
  Ref.\cite{BI}, for "MU[67]" see Ref.\cite{MU} and for "PE[70]" see
  Ref.\cite{PE}. The drawn curves represent various GWU/SAID solutions discussed
  in the text.  
}
\label{fig2}
\end{figure*}

\section{Comparison to existing partial-wave solutions}

\begin{figure*}
\centering
\includegraphics[width=0.79\columnwidth]{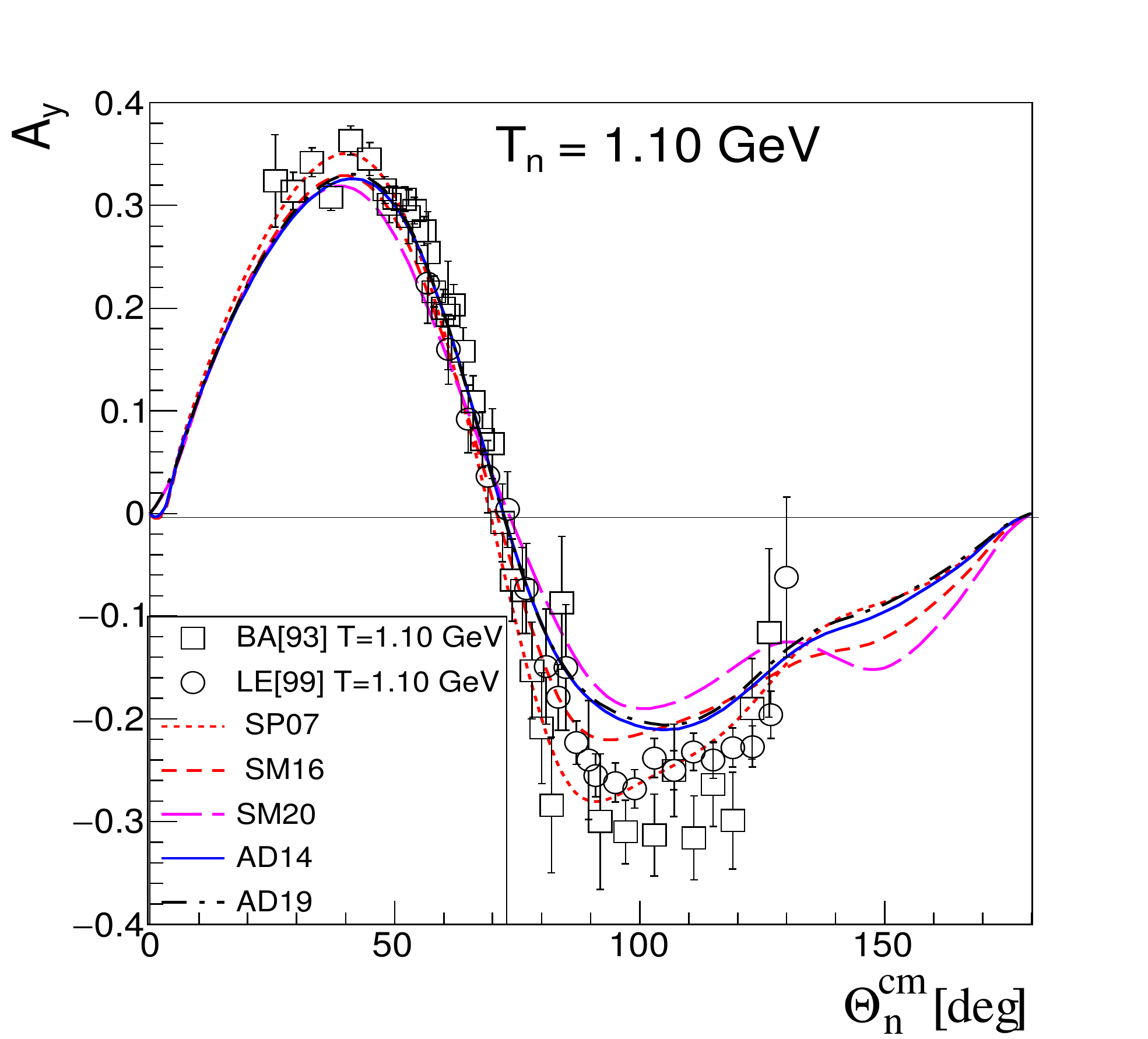}
\includegraphics[width=0.79\columnwidth]{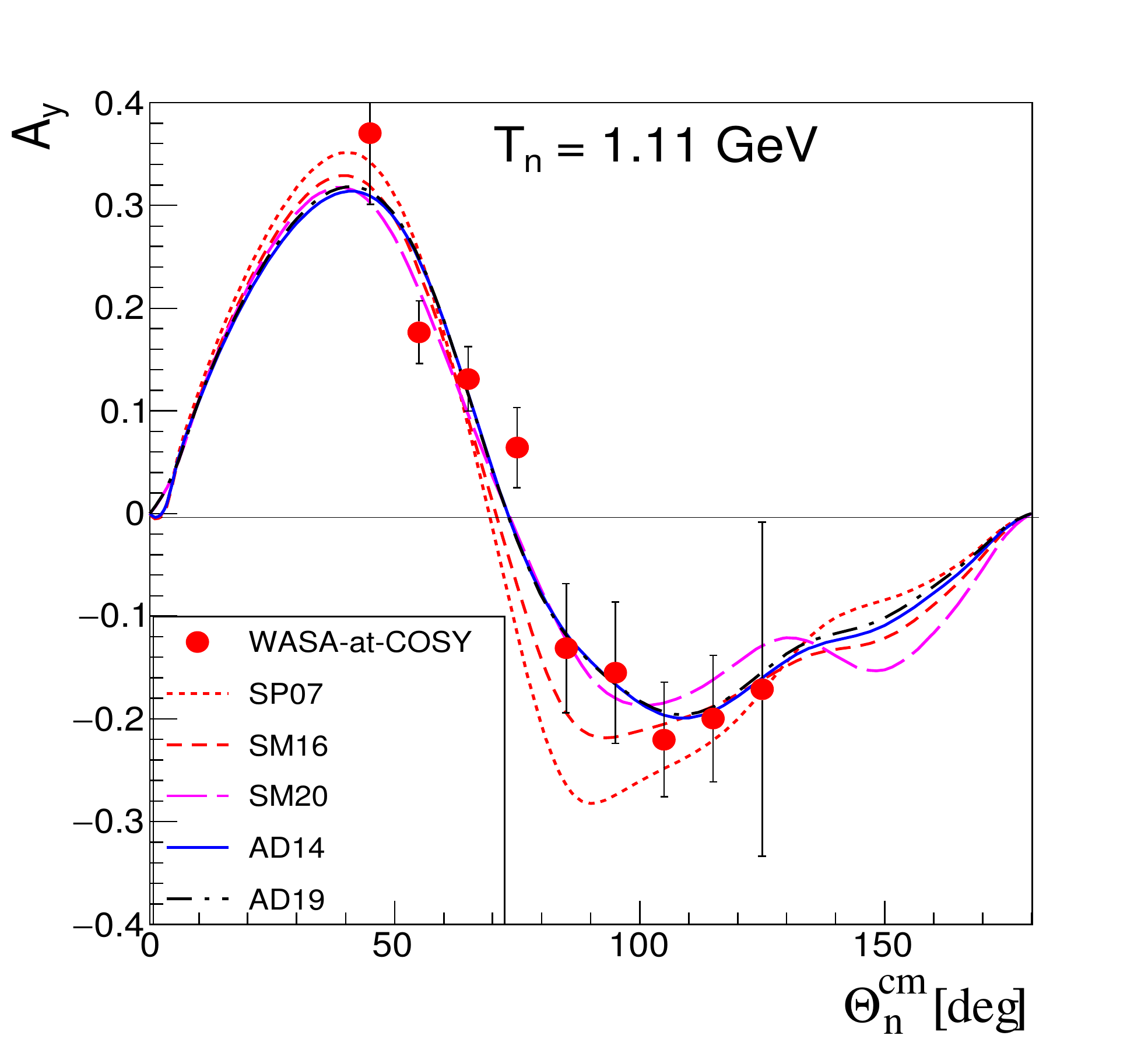}
\includegraphics[width=0.79\columnwidth]{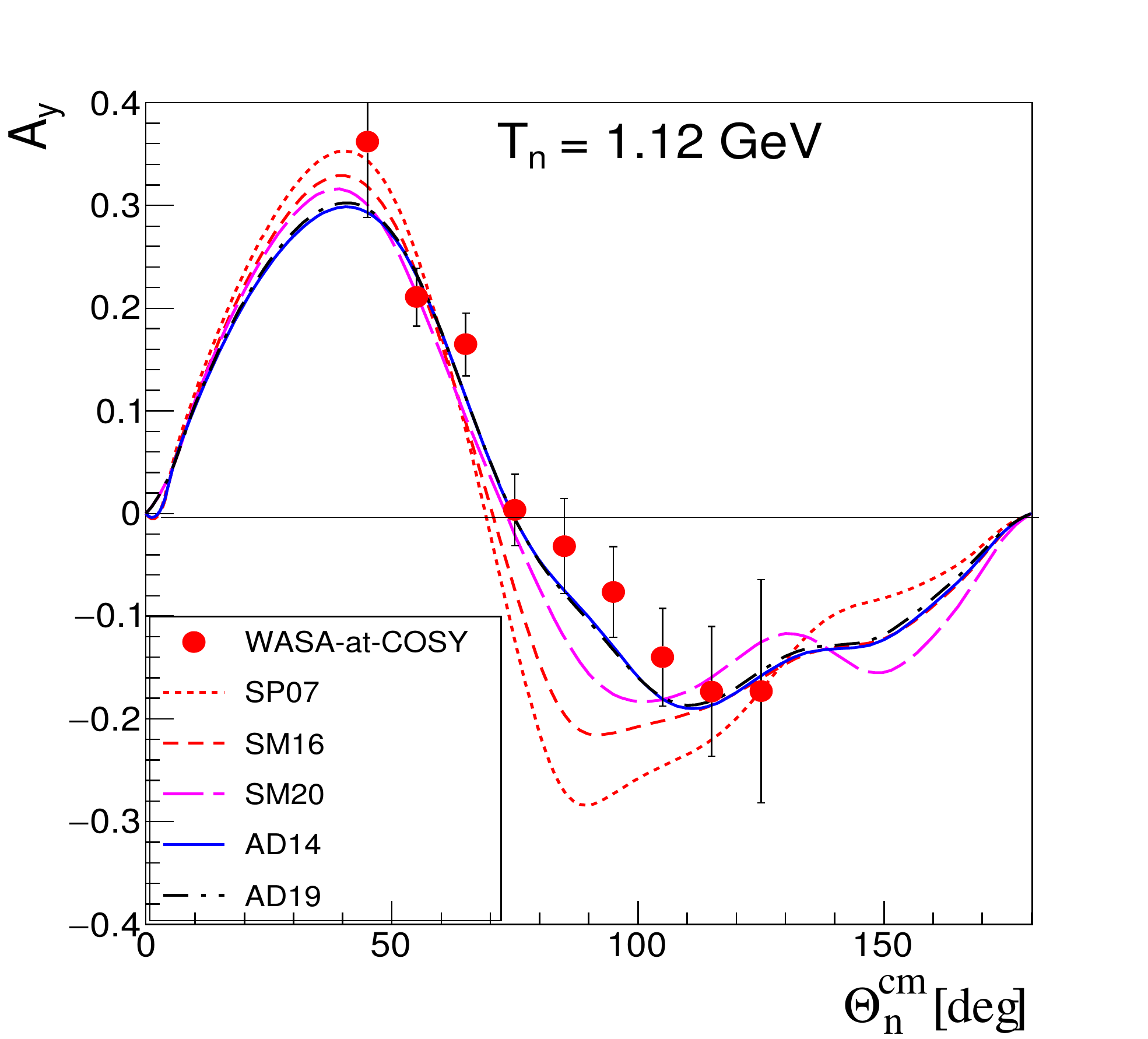}
\includegraphics[width=0.79\columnwidth]{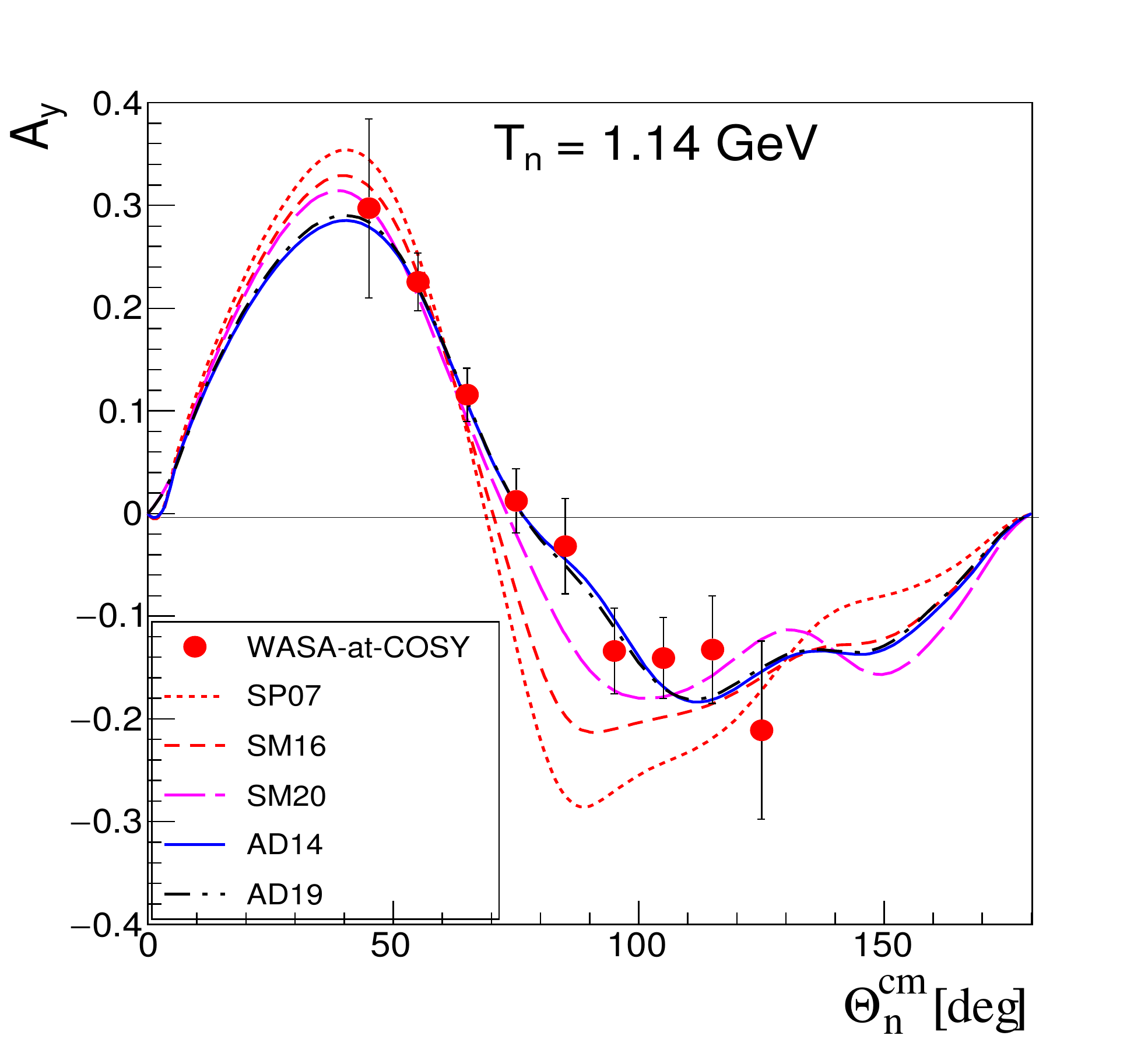}
\includegraphics[width=0.79\columnwidth]{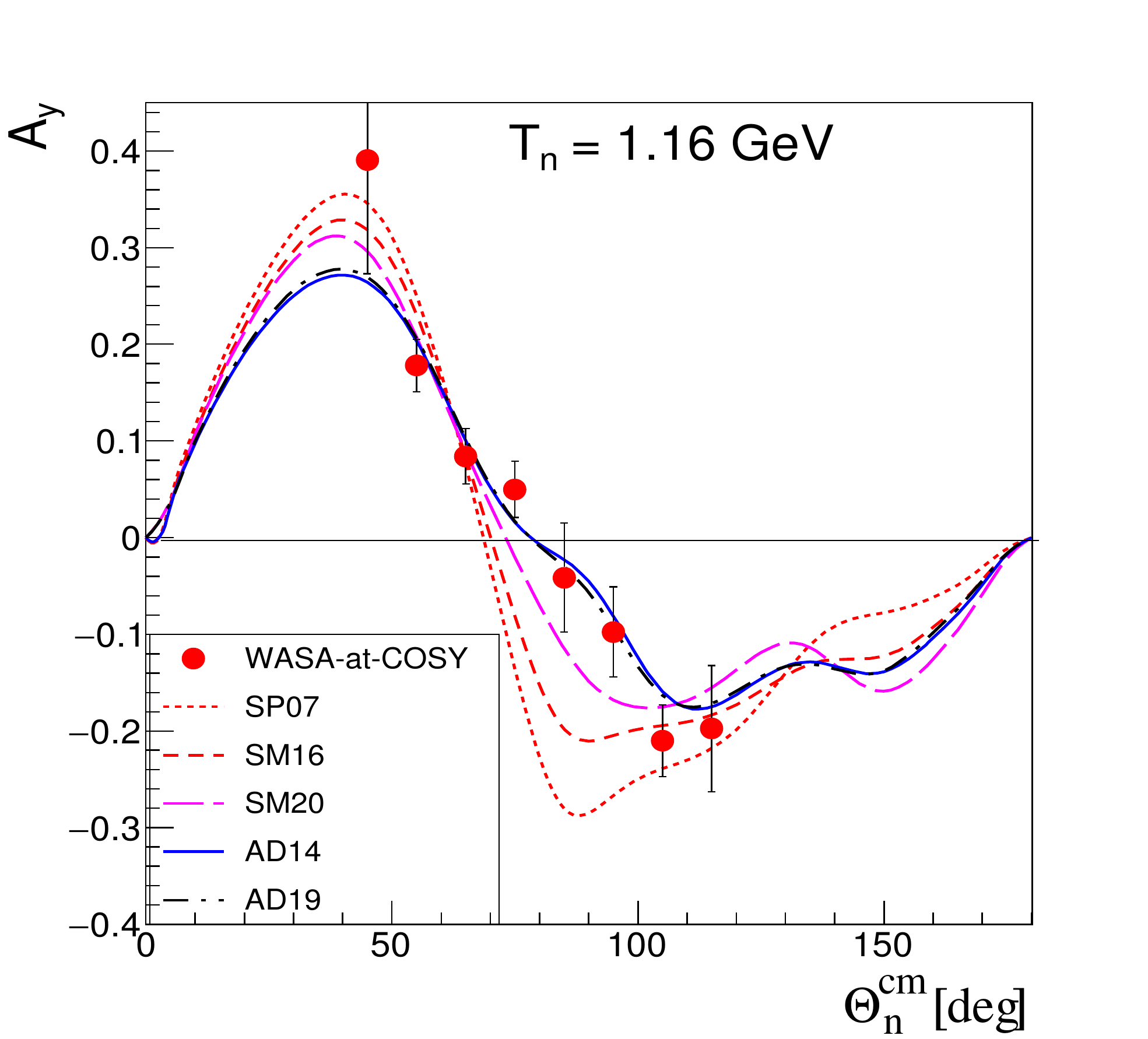}
\includegraphics[width=0.79\columnwidth]{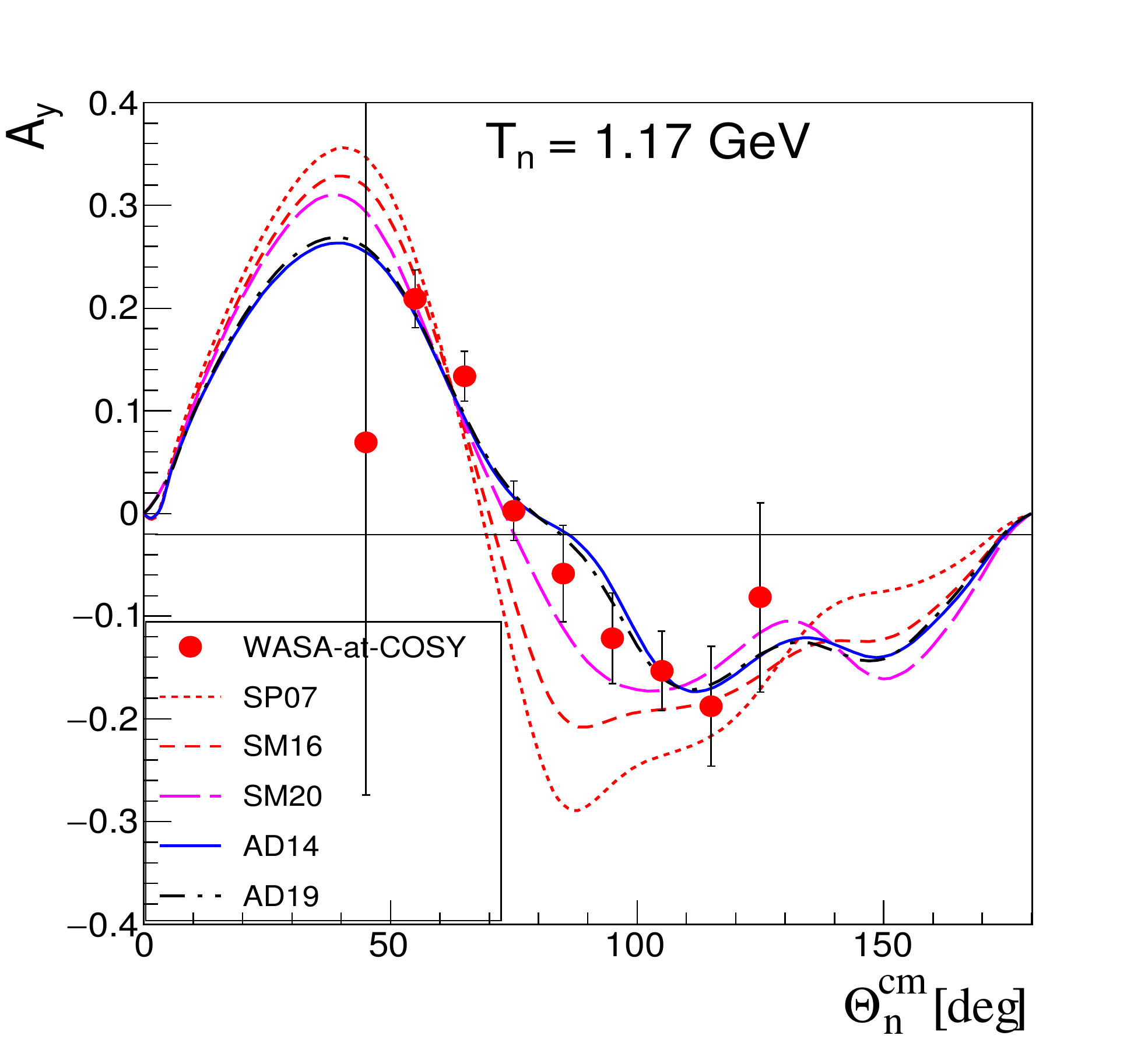}
\includegraphics[width=0.79\columnwidth]{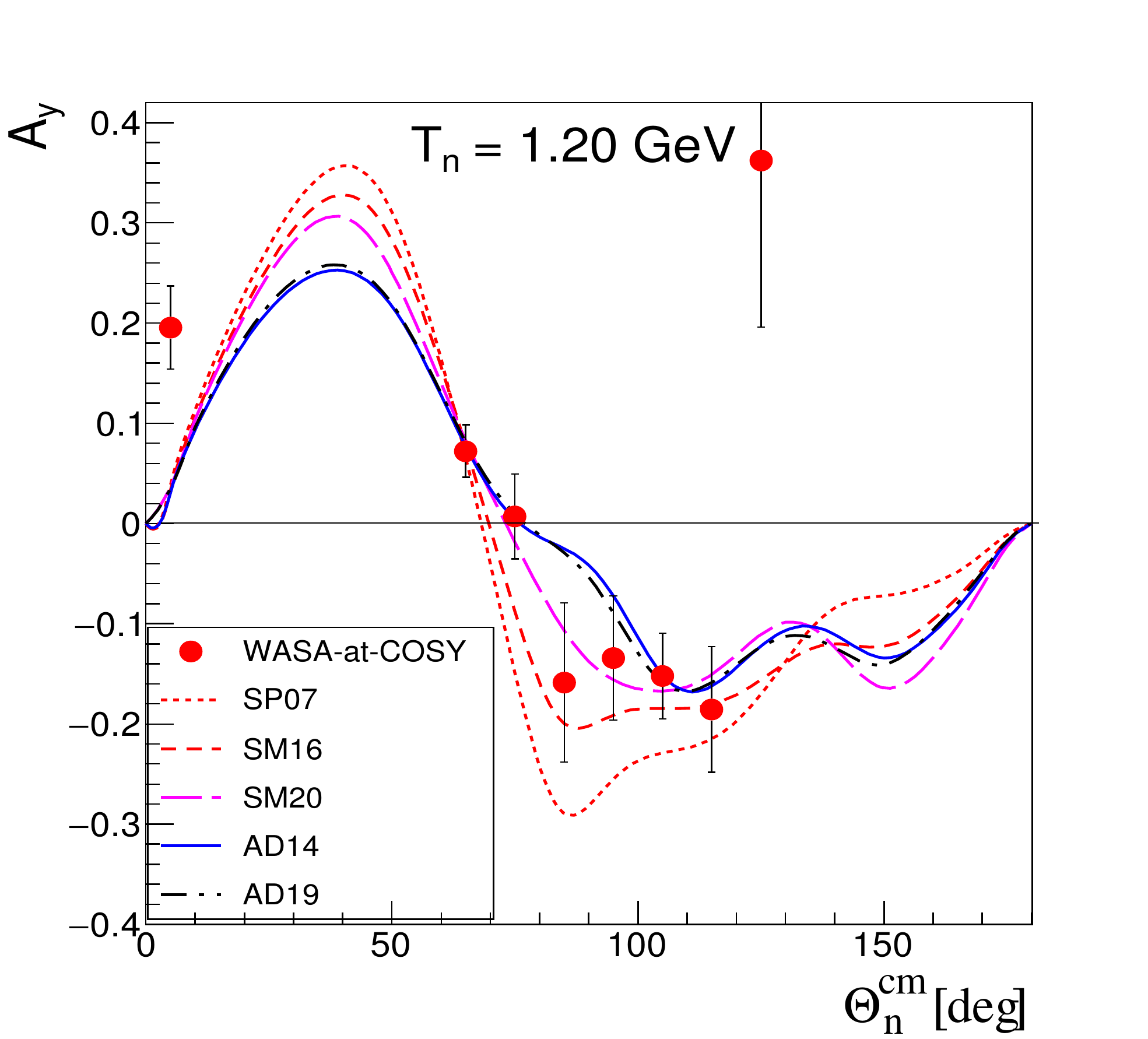}
\includegraphics[width=0.79\columnwidth]{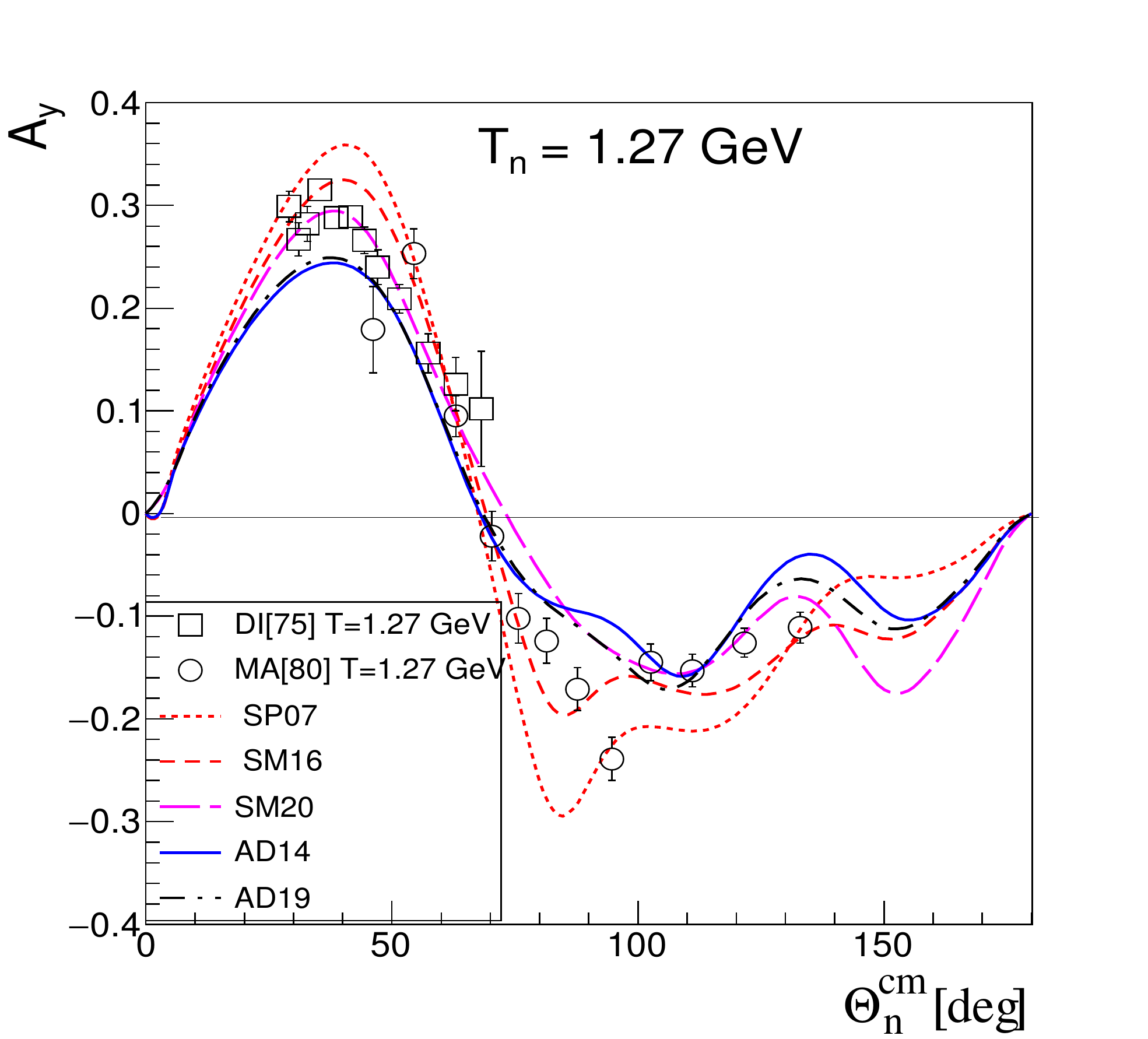}
\caption{\small  Same as Fig.~\ref{fig2}, but for angular
  distributions of the analyzing power.The full symbols denote results from
  WASA-at-COSY \cite{np,npfull}. Open symbols refer to previous
  measurements: for "BA[93]" see Ref.\cite{BA}, for "LE[99]" see
  Ref.\cite{LE}, for "DI[75]" see Ref.\cite{DI} and for "MA[80]" see
  Ref.\cite{MA}.  
}
\label{fig3}
\end{figure*}

In Fig.~\ref{fig2} the data are compared to recently obtained GWU/SAID
partial-wave solutions. The dotted lines resemble the solution SP07
\cite{SP07}, which is based on $NN$ scattering data available until 2007. The
dashed curves represent the solution SM16 \cite{SM16}, which in addition is
based on forward-angle $pp$-scattering data from COSY-ANKE. Both these
solutions do not include the pole of $d^*(2380)$ and hence do not provide a
good description for the $np$ analyzing power data \cite{np,npfull} measured
by WASA-at-COSY in the region of the $d^*(2380)$ resonance as depicted in
Fig.~\ref{fig3}. These analyzing power data,
however, were included in the solution AD14 resulting in a resonance pole for
$d^*(2380)$ in the coupled $^3D_3-^3G_3$ partial waves
\cite{np,npfull,AD14}. This solution is denoted by solid lines in
Figs.~\ref{fig2} and~\ref{fig3}. 

Whereas the SP07 and SM16 solutions give very similar results and provide only
a qualitative description of the differential cross section data, the AD14
solution succeeds to describe these data already quantitatively --- with the
exception of the Birmingham data at $T_p$ = 0.991 GeV \cite{MU} and the
backangle data of Bizard et al. at $T_p$ = 1.252 GeV \cite{BI}.

\begin{figure*}
\centering
\includegraphics[width=0.99\columnwidth]{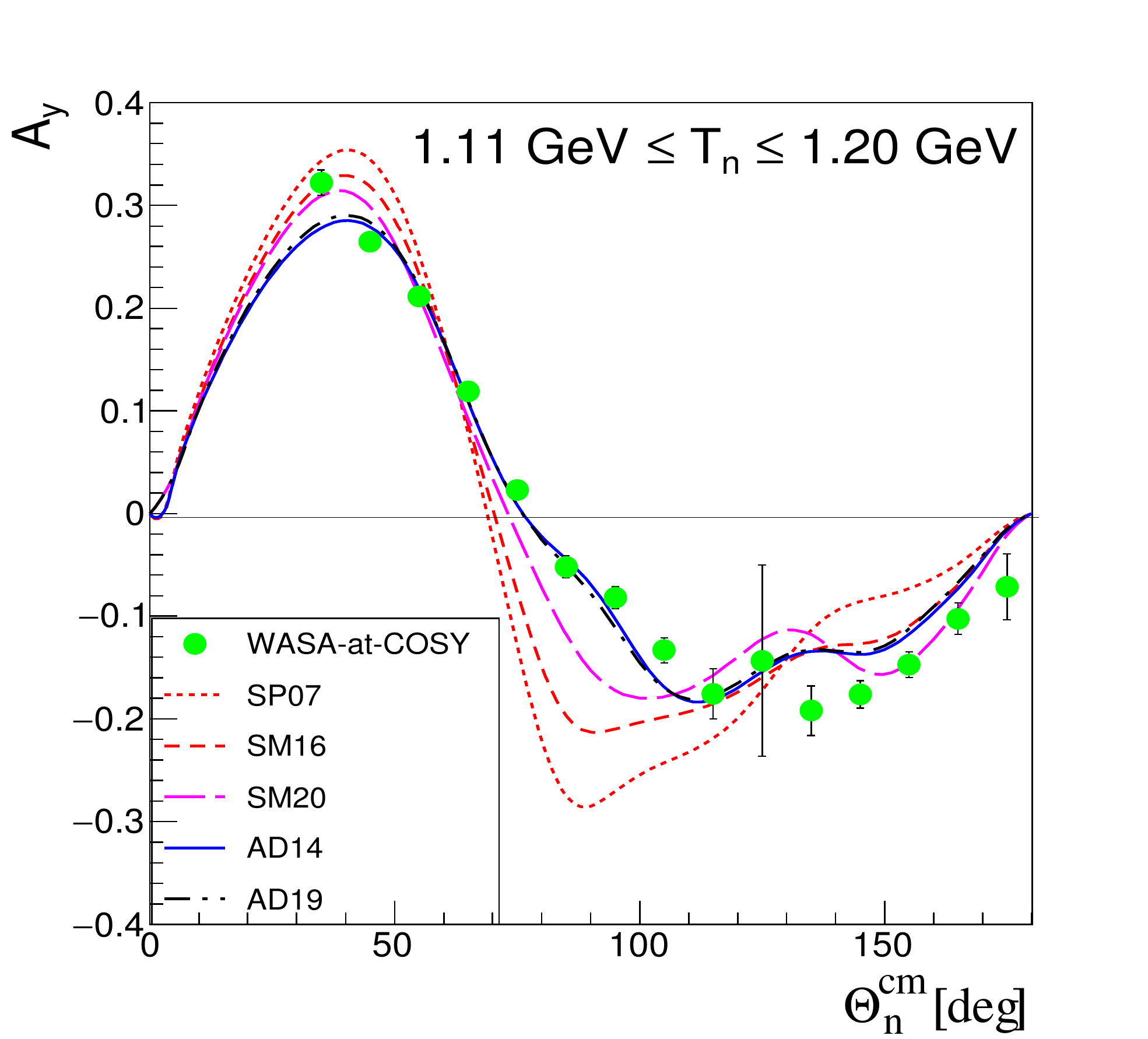}
\includegraphics[width=0.99\columnwidth]{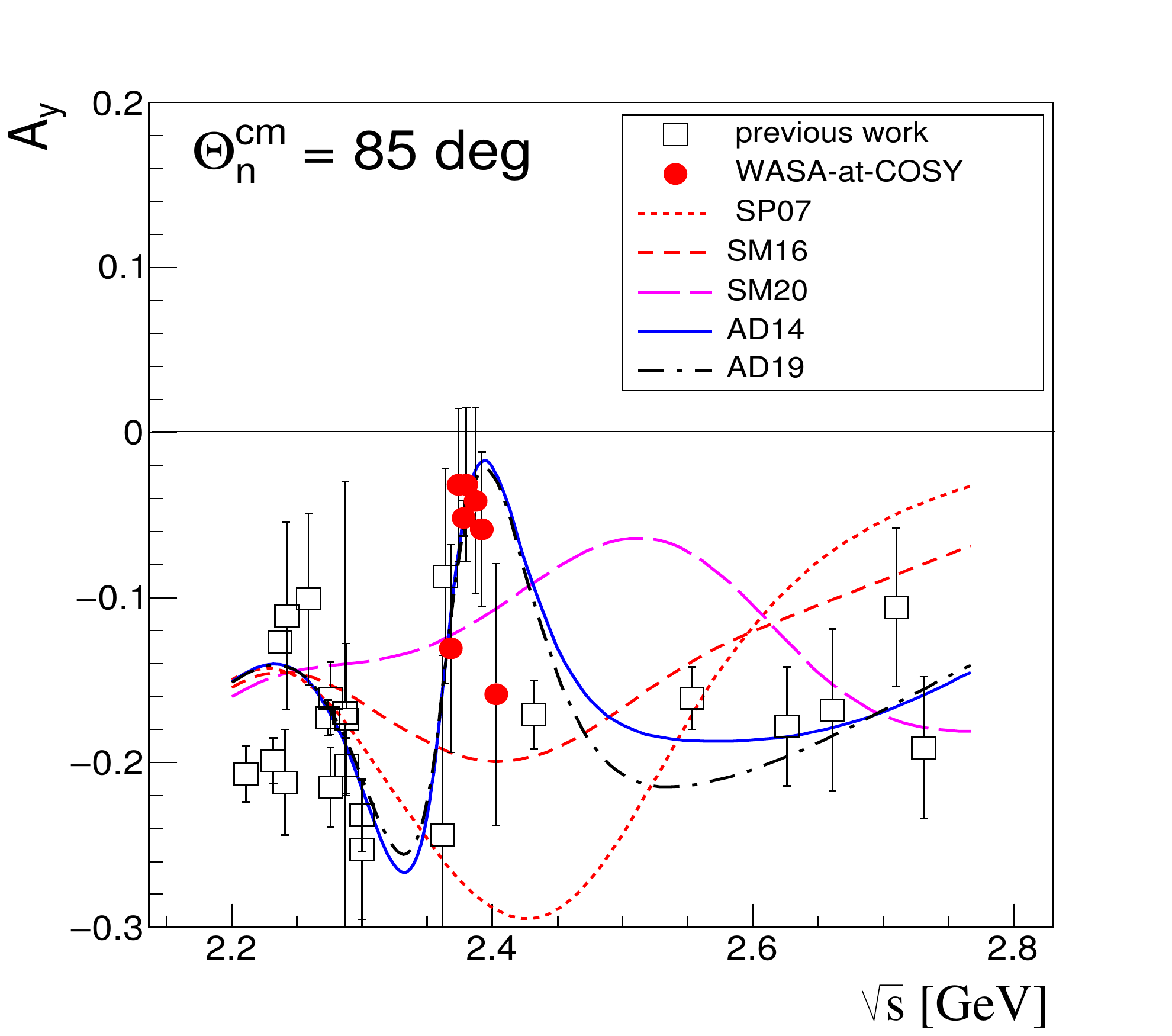}
\caption{\small  Same as Fig.~\ref{fig3}, but the for angular
  distribution of the analyzing power averaged over the resonance region (1.11
  GeV $\leq T_n \leq$ 1.20 GeV) (left) and the energy dependence of the
  analyzing power at $\Theta_n^{cm}$ = 85$^\circ$ (right). Filled circles
  denote the WASA results \cite{np,npfull}, the open squares show previous
  experiments \cite{BA,LE,DI,MA,new,arn,bal1,mcn,gla}. 
}
\label{fig4}
\end{figure*}

\section{Partial-wave analyses including the new cross section data}
 
The WASA-at-COSY cross section data were also included in the GWU/SAID
data base for a new partial-wave analysis. Since the AD14 solution gives
already a quantitative description for the new cross section data, it is of no
surprise that the inclusion of these data in the new partial-wave analysis has
no big impact and hence the resulting new solution AD19 is only marginally
different from the AD14 solution.
The small
differences between AD14 and AD19 solutions may serve as a measure of the
uncertainties in these solutions. 

In order to check the uniqueness of these solutions many fits were carried out
with varying initial weights for different data sets and other starting
conditions. In this attempt, indeed a solution SM20 was found, which comes
closer to the analyzing power data than the previous solutions SP07 and SM16 did
---  as depicted in Fig.~\ref{fig3}. However this
solution does much worse than AD14 and AD19 for the differential cross
sections--- see Figs.~\ref{fig2}, where SM20 appears to be very close to the
SP07 and SM16 results.

In order to investigate the SM20 solution in some more detail, we compare the
various GWU/SAID solutions in Fig.~\ref{fig4} with the WASA high-statistics
data for the angular distribution of the analyzing power in the $d^*(2380)$
region (left panel) as well as with the energy dependence of the analyzing
power near 90$^\circ$ (right panel). As pointed out in Refs.~\cite{np,npfull},
the contribution of $d^*(2380)$ in the analyzing power is proportinal to the
associated Legendre polynomial $P_3^1(cos\Theta_n^{cm})$. Therefore the
resonance effect is at maximum in the 90$^\circ$ region. Due to the richer
data base at 85$^\circ$ we prefer to show the energy dependence not for
exactly  90$^\circ$, but for  85$^\circ$  on the right panel in
Fig.~\ref{fig4}. 

The WASA high-statistics data shown in the left panel of Fig.~\ref{fig4} were
obtained by not accounting for the spectator momentum. Thus these data
represent a weighted average  over the measured interval $\sqrt s =$ 2.37 -
2.40 GeV ($T_n =$ 1.11 - 1.20 GeV) with a centroid at $\sqrt s =$ 2.38 GeV
--- see Fig.~1 in Ref.~\cite{np} and Fig.~4 in Ref.~\cite{npfull},
respectively. We see that for the various partial-wave solutions the most
critical angular region is around 90$^\circ$, {\it i.e.} exactly the region which is most sensitive to the $d^*(2380)$ resonance. Whereas AD14 and AD19 solutions
reproduce the experimental data very well in this region, the other solutions
miss the data there. Most striking is the failure of SP07. SM16 does a bit
better and SM20 comes still closer, but nevertheless fails quantitatively in
this angular region.

In the right panel of Fig.~\ref{fig4} we explore the energy dependence of the
analyzing power in this angular region. There the data exhibit a pronounced
pattern resembling the interference of a narrow resonance with the
background. The solutions AD14 and AD19 are able to reproduce this pattern
quantitatively, though the data suggest a somewhat narrower resonance pattern
at the high-energy side\footnote{This would be in accord with a slightly
  narrower resonance width as it is observed in fact in the $NN \to NN\pi\pi$
  channels \cite{MB}.}. 
The SP07 solution predicts a smoothly curved energy dependence, which
is far off the data, whereas the SM16 and SM20 solutions exhibit a somewhat
flatter energy dependence coming thus  closer to the data on average, but still
severely miss the resonance structure in the energy region of $d^*(2380)$.

We conclude that the solutions SP07, SM16 and SM20 all fail in a quantitative
description of both cross section and analyzing power data in the energy
region of $d^*(2380)$, whereas the solutions AD14 and AD19 can account
quantitatively for all experimental data.

\begin{figure*}
\centering

\includegraphics[width=0.69\columnwidth,angle=90]{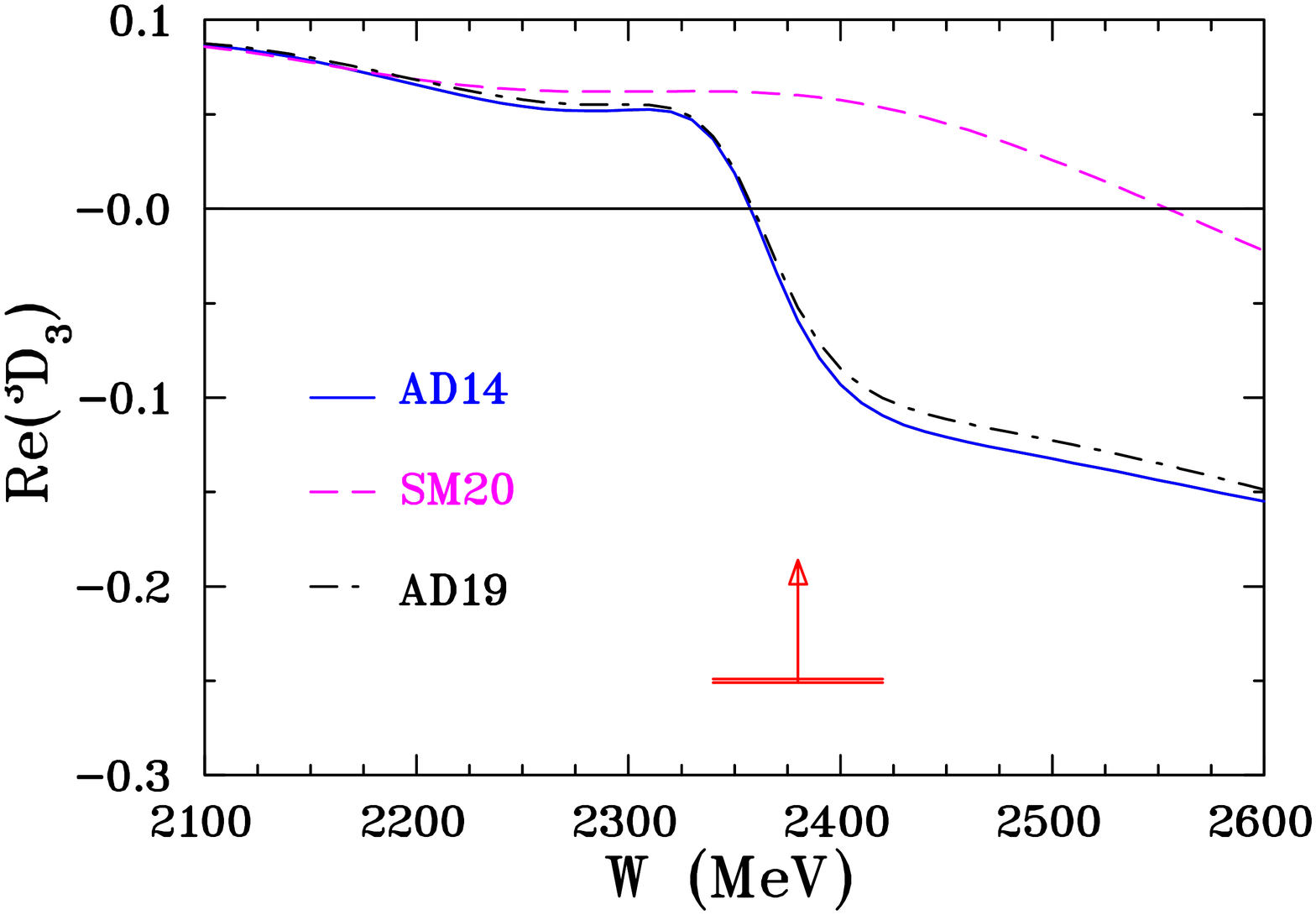}
\includegraphics[width=0.69\columnwidth,angle=90]{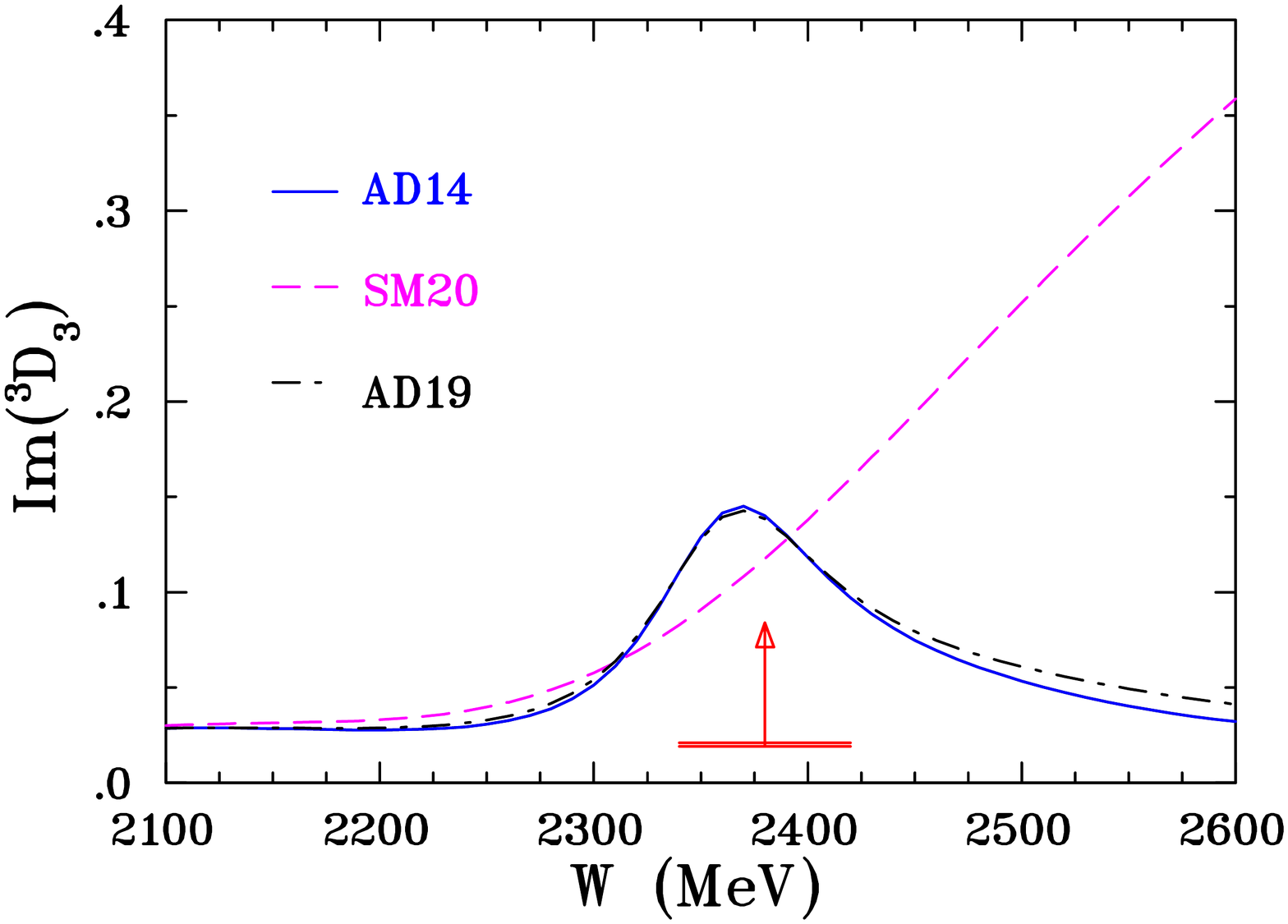}
\includegraphics[width=0.69\columnwidth,angle=90]{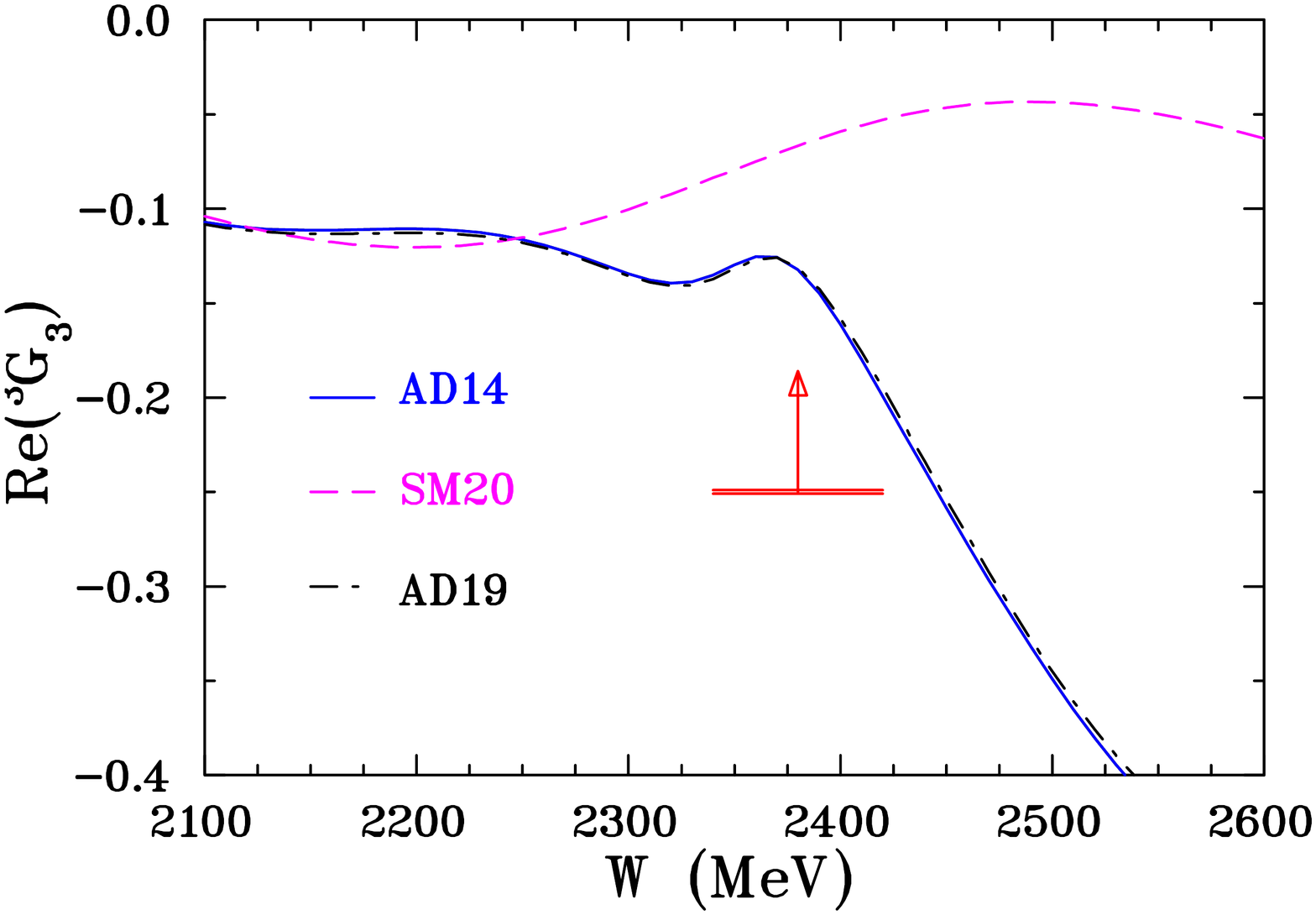}
\includegraphics[width=0.69\columnwidth,angle=90]{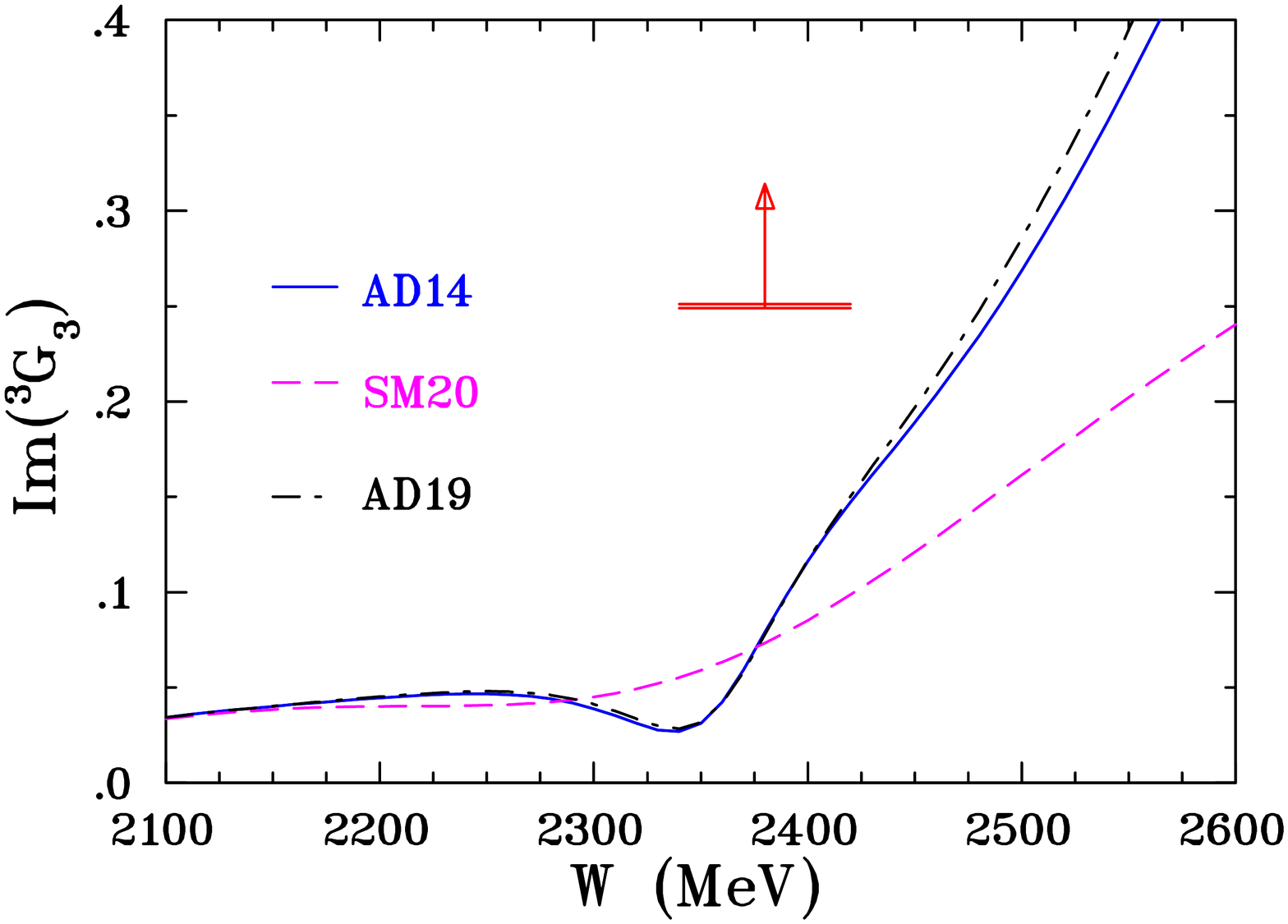}
\includegraphics[width=0.69\columnwidth,angle=90]{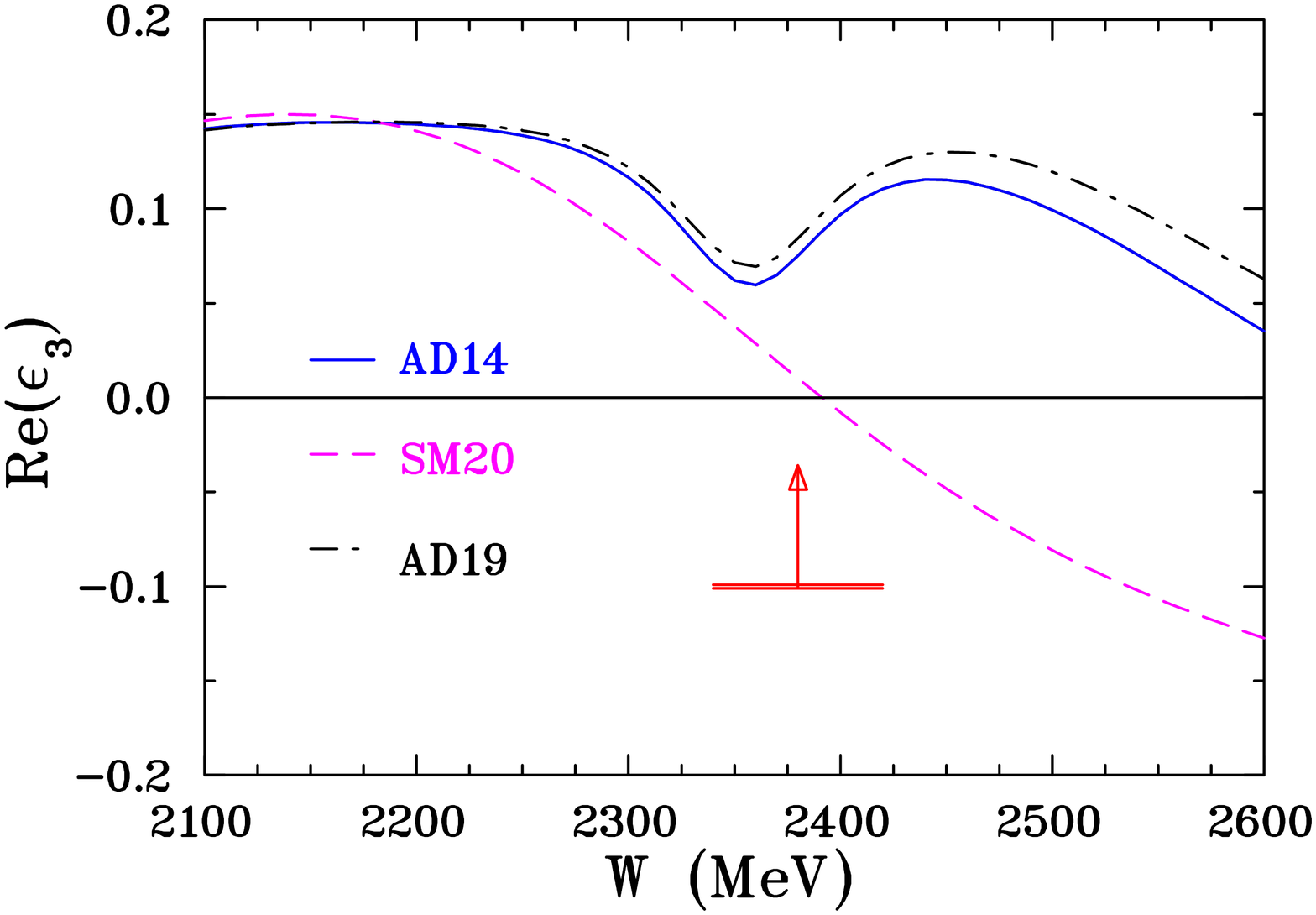}
\includegraphics[width=0.69\columnwidth,angle=90]{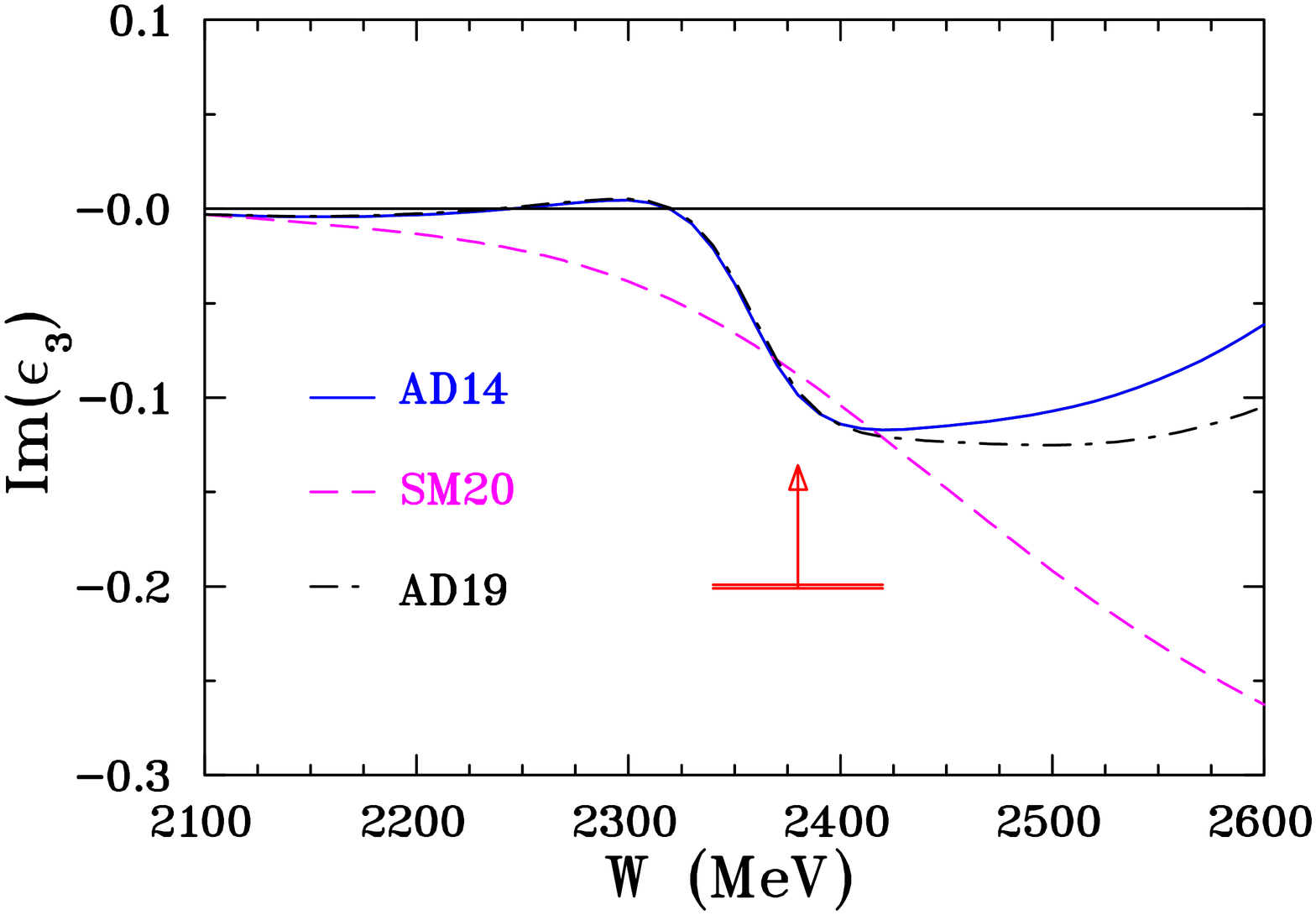}
\caption{\small Real (left) and imaginary (right) parts of the $^3D_3$ (top),
  $^3G_3$ (middle) partial wave amplitudes and the mixing term
  $\epsilon_3$ (bottom) for the GWU/SAID solutions SM20 (magenta, dashed),
  AD14 (blue, solid) and AD19 (black, dash-dotted).Vertical arrows with
  horizontal bar indicate the location of mass and width of $d^*(2380)$.   
}
\label{FigGWU}
\end{figure*}

In Fig.~\ref{FigGWU} we plot the $^3D_3$ and $^3G_3$ partial-wave amplitudes as
well as their mixing term $\epsilon_3$ in dependence of the center-of-mass
energy $W = \sqrt s$ for the solutions AD14 (solid), AD19 (dash-dotted) and
SM20 (dashed). The amplitudes for the AD14 and AD19 solutions are very close
and differ slightly only at high energies. Both solutions exhibit a clear
resonance structure in the $d^*(2380)$ region in both real and imaginary parts
of all three amplitudes. In contrast, the SM20 solution exhibits only a very
smooth energy dependence without indication of any resonance.

\begin{figure}
\centering
\includegraphics[width=0.79\columnwidth]{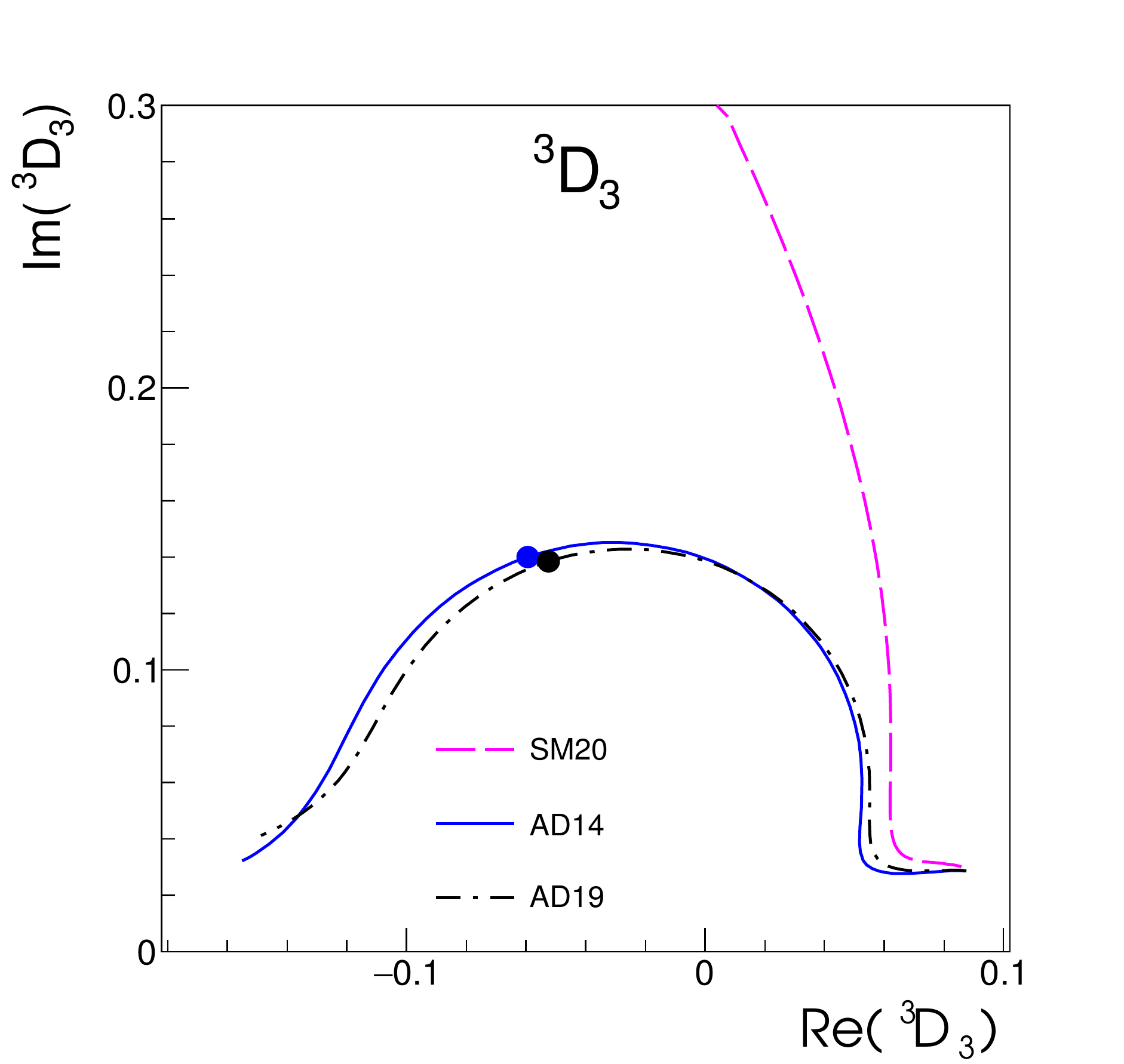}
\includegraphics[width=0.79\columnwidth]{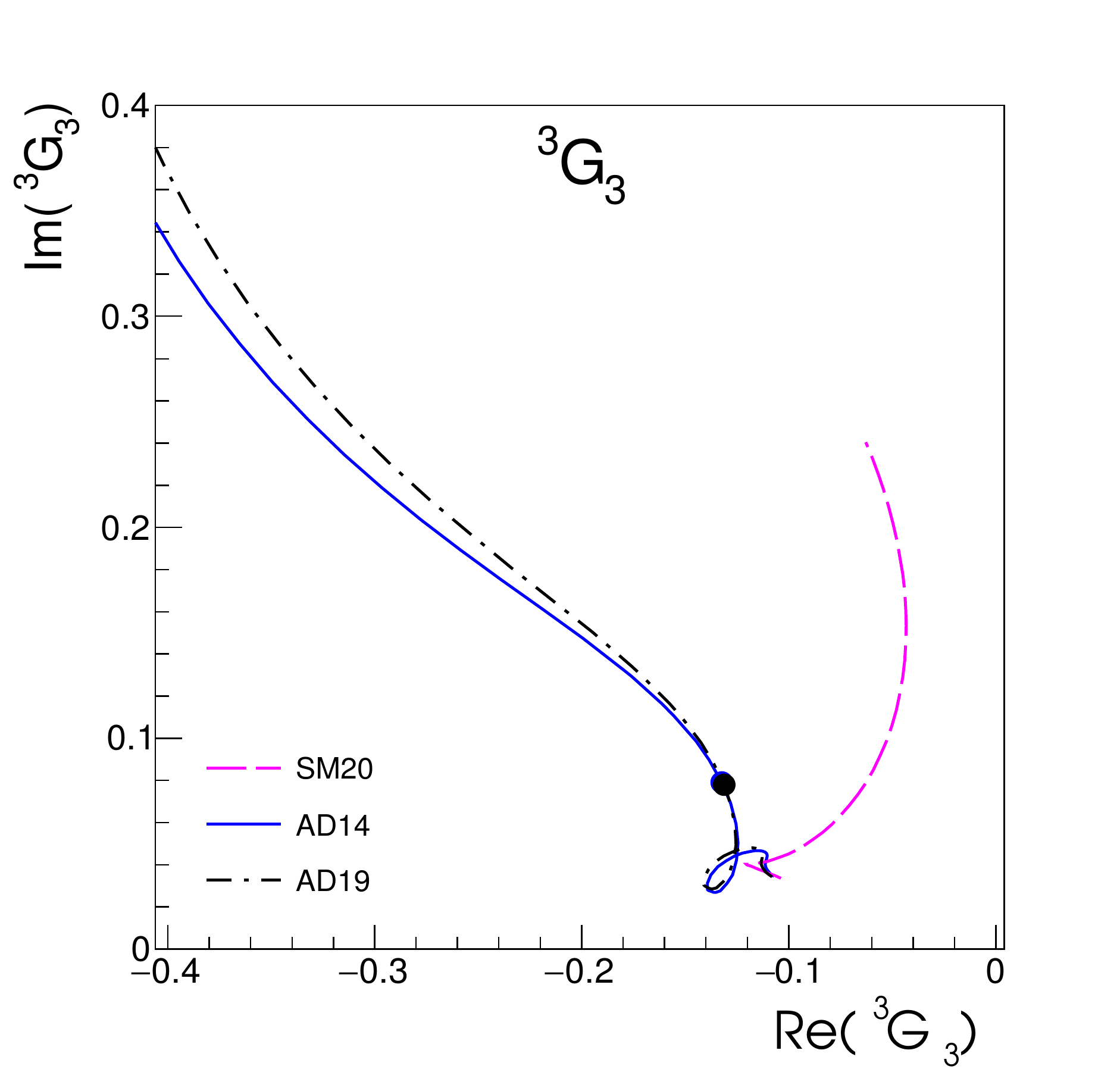}
\includegraphics[width=0.79\columnwidth]{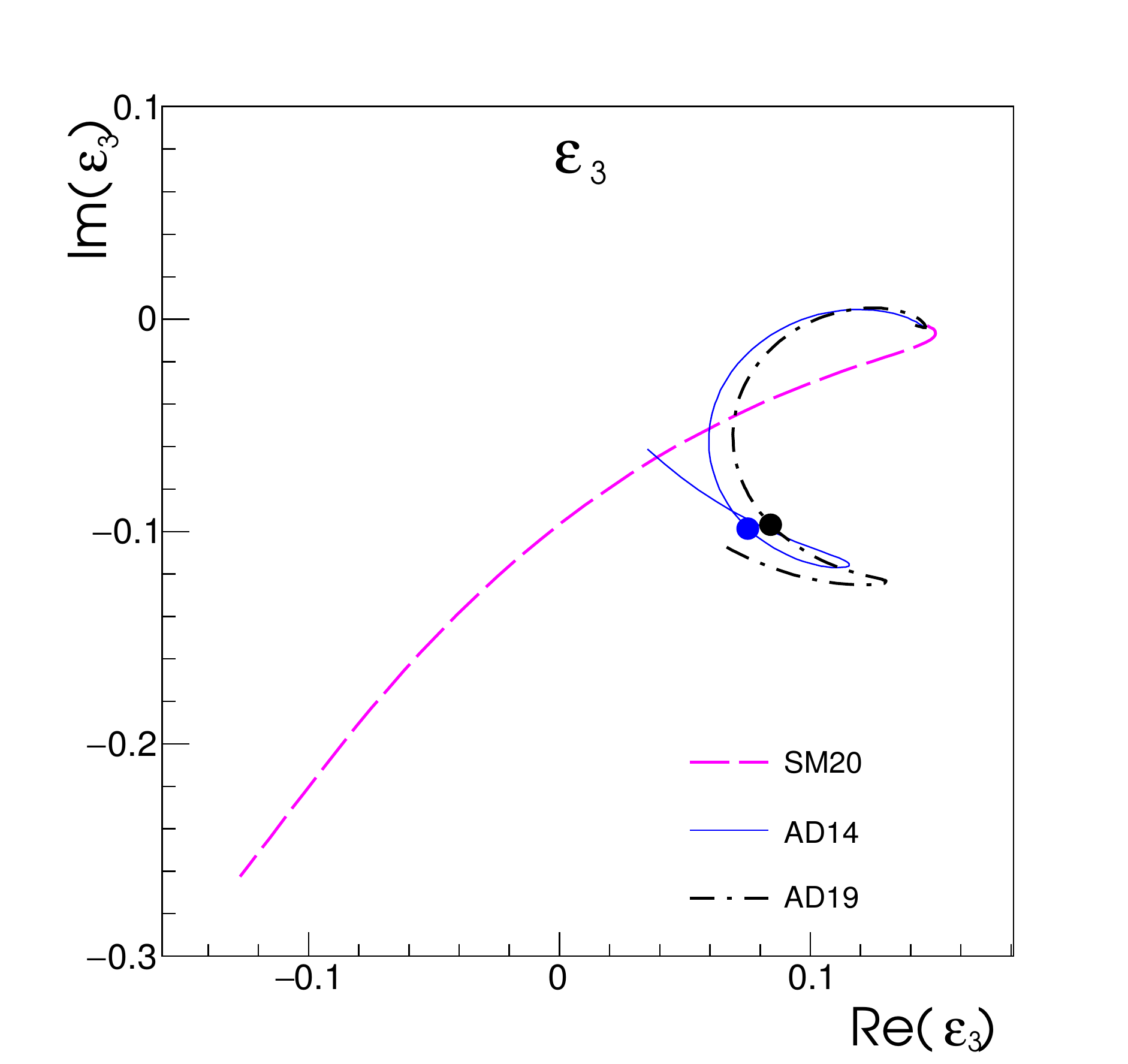}
\caption{\small Argand diagrams for the $^3D_3$ (top) and
  $^3G_3$ (middle) partial wave amplitudes and their mixing term $\epsilon_3$
  (bottom) for the GWU/SAID solutions SM20 (magenta, dashed), AD14 (blue,
  solid) and AD19 (black, dash-dotted). The thick dots display the $d^*(2380)$
  pole position
}
\label{fig6}
\end{figure}

In Fig.~\ref{fig6} we plot the Argand diagrams for the $^3D_3$ (top) and
$^3G_3$ (middle) partial waves and their mixing amplitude $\epsilon_3$
(bottom) for the GWU/SAID solutions SM20 (magenta, dashed), AD14 (blue, solid)
and AD19 (black, dash-dotted). Whereas the SM20 solution shows no obvious
looping in these diagrams, {\it i.e.} no sign of a pole, the solutions
AD14 and AD19, which nearly coincide, do exhibit pronounced
loops in accordance with the presence of the $d^*(2380)$ pole.

\section{Partial-Wave Analysis and Data Interpretation}

In the following 
the search for poles presented in ref.~\cite{npfull}, based on analyzing
Argand diagrams and the speed plot, is improved. Namely, a looping in the
Argand diagram is
in the mathematical sense only a necessary condition for the existence of a 
pole, but not yet a sufficient one. This simply means that, if a function has a
pole, 
it must produce a backward looping, however a backward looping found in
Argand diagrams  can be produced also by other effects. {\ it E.g.}, a backward
looping in the Argand diagram can be produced by branch-points caused by
channel openings  --- in our case by the nearby $NN^*(1440)$
threshold. 
To prove definitely 
that we indeed have a pole we need a more stringent method. Therefore, instead  
 of analyzing Argand diagrams of $^3D_3$, $^3G_3$, and mixing term
 $\epsilon_3$ we introduce the trace of the  $^3D_3$ - $^3G_3$  matrix, and
 instead of quantifying the $^3D_3$ partial wave with the speed plot technique
 we quantify the whole trace with the Laurent+Pietarinen(L+P) expansion. 
\\ \\ \noindent
The coupled partial waves in question are creating the following
$I(J^P)=0(3^+)$ mixing matrix. 
\begin{eqnarray}
T=
\begin{bmatrix}
&^3D_3 & \epsilon_3  \\
\epsilon_3 &  ^3G_3
\end{bmatrix}
\end{eqnarray}

Without the loss of generality this matrix can be given by its Laurent
decomposition in its area of convergence: 
\begin{eqnarray}
T  & = &
\begin{bmatrix}
\frac{a_{11} + \di \, b_{11}}{Den} + B_{11} & \frac{a_{12} + \di \, b_{12}}{Den}+ B_{12}    \\ \\  \nonumber
\frac{a_{12} + \di \, b_{12}}{Den} + B_{12}  &  \frac{a_{22} + \di \, b_{22}}{Den} + B_{22}
\end{bmatrix}
\\ \nonumber \\
& & Den  =  M - W - \di \, \Gamma
\end{eqnarray}
The $^3D_3$, $^3G_3$ partial waves and the mixing term $\epsilon_3$ are given in Fig.~\ref{FigGWU} for
the  GWU/SAID solutions SM20 (magenta, dashed), AD14 (blue,
solid) and AD19 (black, dash-dotted).

\begin{figure}
\centering
\includegraphics[width=0.99\columnwidth]{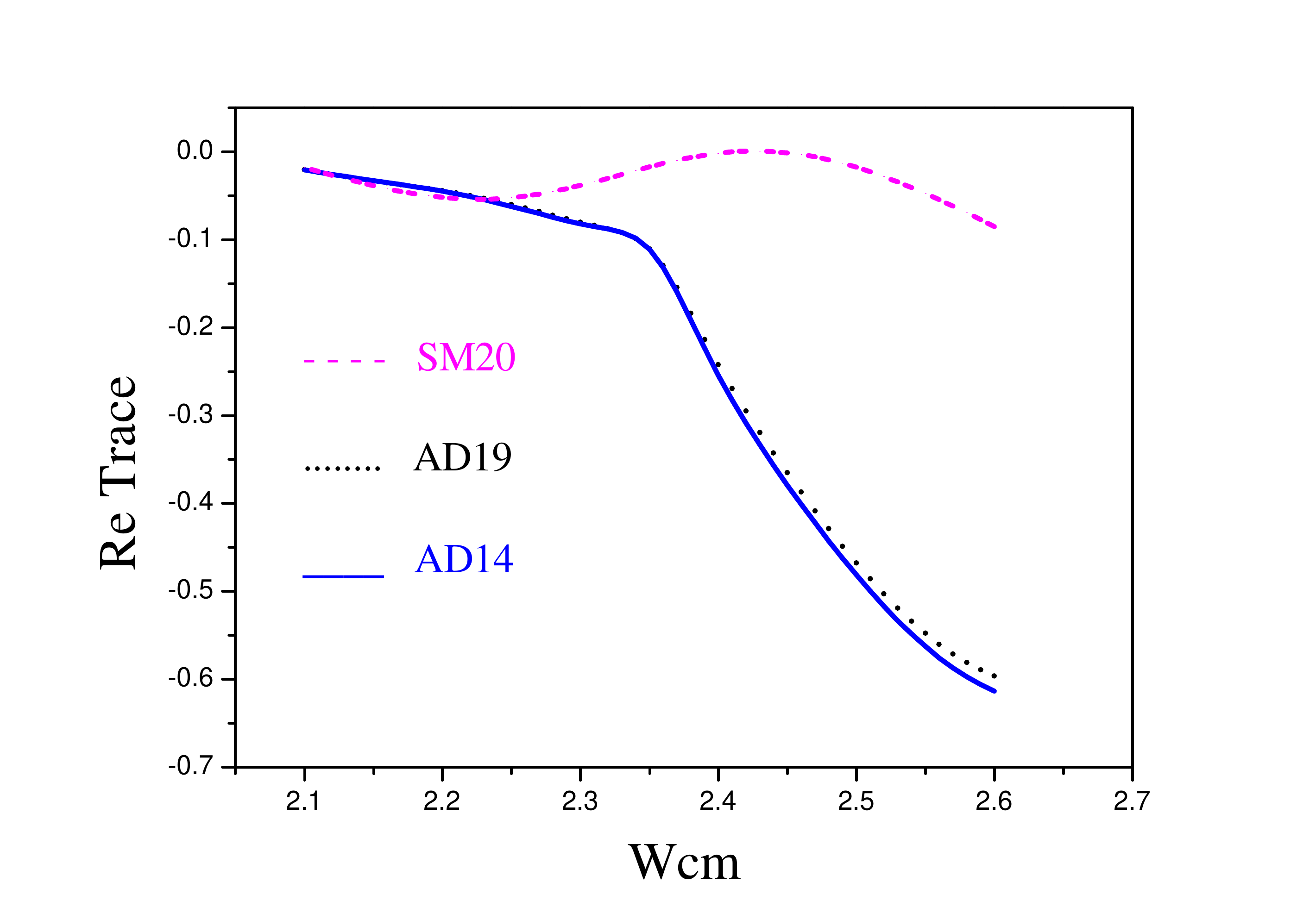}
\includegraphics[width=0.99\columnwidth]{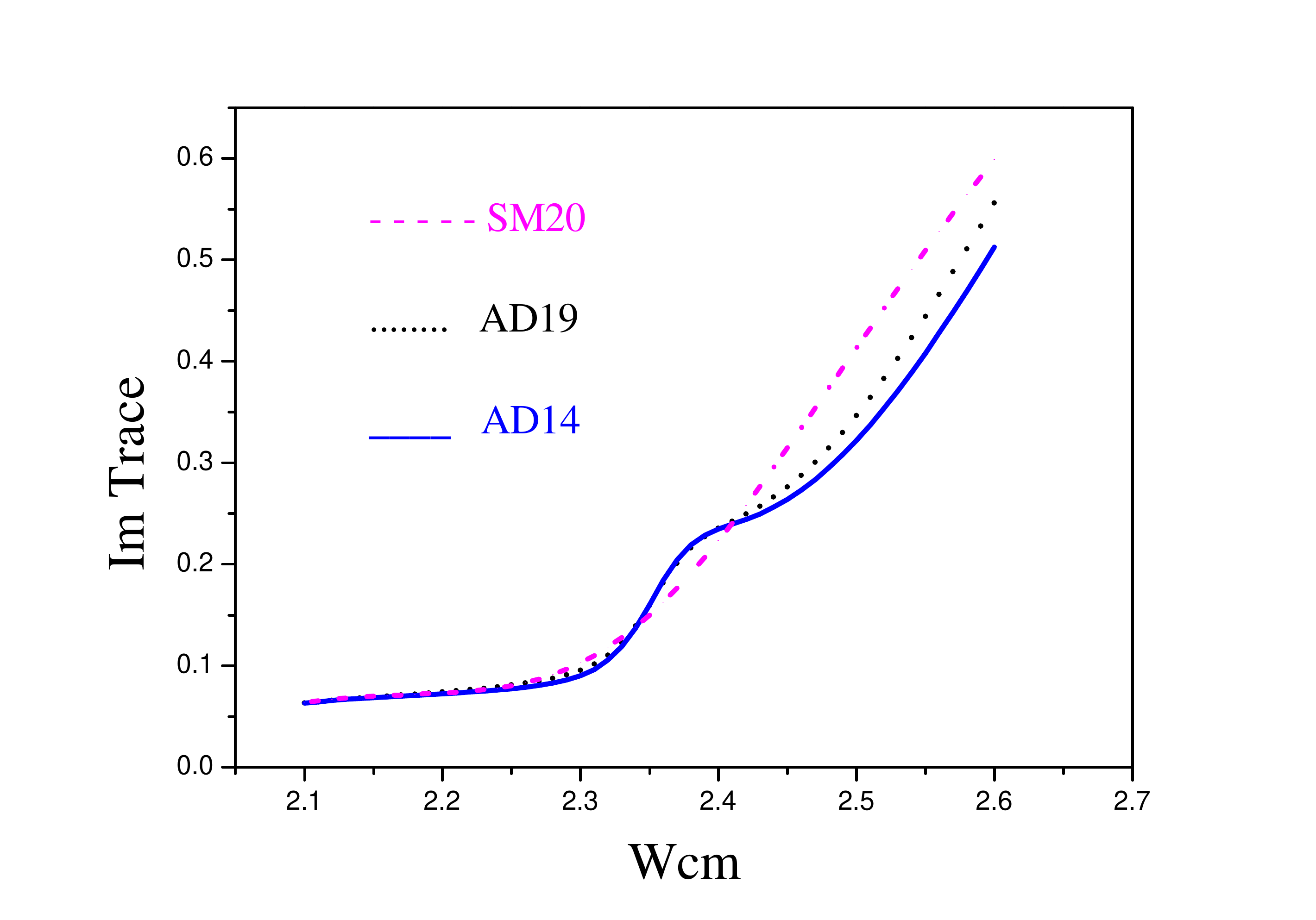}
\caption{\small The Trace of all three   GWU/SAID partial-wave solutions SM20
  (magenta, dashed), AD14 (blue, solid) and AD19 (black, dotted) solutions. } 
\label{Figtrace}
\end{figure}

 Following the idea presented in ref.~\cite{Ceci2008} we use the
 trace\footnote{Trace of the matrix is defined as the sum of diagonal matrix
   elements, and due to its commutativity as \mbox{Trace(${\bf A} \cdot {\bf
       B}$ )= Trace($ {\bf B} \cdot {\bf A}$)} it is identical for all
   matrices obtained from the original matrix by $U^{-1} \cdot {\bf A}
   \cdot U$; hence for the diagonal one too.} 
of $I(J^P)=0(3^+)$ mixing matrix:
\begin{eqnarray}
{\rm Trace [T] } & = & \frac{ (a_{11} +  a_{22}) + \di \, (\, b_{11}  + \, b_{22})}{Den}+ \left( B_{11}+ B_{22} \right). \nonumber  \\
\end{eqnarray}
As it has been shown in that reference, structures which are buried under
notable background in individual matrix elements pop out once the trace of the
matrix is done. In Fig.~\ref{Figtrace} we show the trace of all three
GWU/SAID solutions.

The quantitative evaluation of the pole parameters of the Trace [T] is done
using the Laurent+Pietarinen (L+P) formalism identical to the way how it has
been done in ref.~\cite{Briscoe2019}. For the convenience of the reader let us
repeat some basic facts. 

The driving concept behind the method is to replace the complexity of solving
an elaborate theoretical model and analytically continuing its solution into
the complex energy plane by using a local power-series representation of partial
wave amplitudes which just exploits analyticity and unitarity. The L+P
approach separates pole and regular part in the form of a Laurent expansion,
and instead of modeling the regular part using some physical model it uses the
conformal-mapping-generated, rapidly converging power series with well defined
analytic properties 
called a Pietarinen expansion to represent it effectively. In other words, the
method replaces the regular part calculated in a model with the simplest
analytic function which has 
correct analytic properties of the analyzed partial wave (multipole), and fits
the given input. In such an approach the model dependence is minimized, and is
reduced to the 
choice of the number and location of L+P branch-points used in the model.

So, we expand the trace in terms of a sum over all poles and
with a Pietarinen series representing the  energy dependent regular (non-pole)
part as: 

\begin{eqnarray}
	{\rm Trace \, [T]} & = & \sum_{i=1}^{k}\frac{\alpha_{-1}^{(i)}}{W - W_i} + B^L(W) .
	\label{eq:LP}
\end{eqnarray}\noindent
Here $W$, $\alpha_{-1}^{(i)}$, and W$_i$ are complex numbers
representing the c.m. energy, residues, and pole positions for the
\textit{i}th pole, respectively, and B$^L$(W) is a regular function
in the whole complex plane. As it has been shown in Ref.~\cite{Svarc:2013sqa}
the generally unknown analytic function
$B(W)$ with branch-points in $x_P$, $x_Q$, and $x_R$ can be expanded into a
power series of Pietarinen functions
as
\begin{eqnarray} \label{L+P}
	B^L(W)& = & \sum_{n=0}^{M}c_nX(W)^n +  \sum_{n=0}^{N}d_nY(W)^n \nonumber \\
	  & + & \sum_{n=0}^{N}e_nZ(W)^n +..., \nonumber \\
	  &  &  X(W =  \frac{\alpha - \sqrt{x_P - W}}{\alpha + \sqrt{x_P - W}} , \nonumber \\
      &  &   Y(W)  =  \frac{\beta - \sqrt{x_Q - W}}{\beta + \sqrt{x_Q - W}} , \nonumber \\
      &  &  Z(W) =  \frac{\gamma - \sqrt{x_R - W}}{\gamma + \sqrt{x_R - W}}
\end{eqnarray}\noindent
where c$_n$, d$_n$, e$_n$ and $\alpha$, $\beta$, $\gamma$ are real
numbers that represent tuning parameters and coefficients of the
Pietarinen functions X(W), Y (W), and Z(W), respectively.  A variable
number of coefficients in the three series of Eq.~(\ref{L+P}) was used,
depending on the structure of the non-pole 
part of each amplitude.

As the nearby energy points of the input partial-wave trace are correlated
through analyticity of the energy dependent partial wave of the GWU/SAID
solutions, the standard error analysis cannot be used as the standardly
defined $\chi^2$ becomes 
extremely small ($\chi^2 << 1$) regardless which error is attributed to the
input. The method used is identical to what has been done in
ref.~\cite{Briscoe2019}, and is based on randomizing the central values of the
 energy dpendent (ED) solution with partial-wave (PW) standard
deviation $\sigma_{PW}$,  
and assigning the error of the source ED point as the error of the randomized
point. In that way we generate 1000 different sets which are analyzed by L+P,
and the error analysis is done in a standard way for non-correlated quantities.

At this point it is important to stress that our central problem is to
establish whether the analyzed GWU/SAID solutions contain a pole or not. The
L+P approach by construction detects resonances in two different ways: either
as a resonance in a two-body process, which manifests itself as a pole on the
real axes, or a resonance in the three body sub-system, which manifests itself
as a complex branch-point. In either of the two cases we encounter a resonance;
however, there is still the matter of identifying its location with the purpose
of its identification. 
 
The difference between the two situations is subtle. If we have a genuine pole
in the two-body system, our real and imaginary parts will show a typical
resonance behaviour, and real branch-points which represent the opening of
two-body channels consisting of two stable particles should be in principle
clearly visible as they  produce sharp cusps in the analyzed
amplitude. However, where the branch-point represents a channel which consists
of a stable particle and a two-body resonant state, this branch-point will
become complex, and the sharp, cusp effect disappears. These two processes are
different, but the method will require much higher precision of the data to
distinguish between the two.   Just by looking at figure Fig.~\ref{Figtrace}
it is clear that our process will strongly depend on the confidence limit of
all obtained GWU/SAID solutions. Namely, in the ideal case, when the
confidence limit is high and the error band is low, the method will be able to
distinguish between the two. However, as soon as error bands become realistic,
the clearly visible peak in the imaginary part will be smeared out, and the
distinction between the two scenarios (genuine pole or complex branch-point)
will be lost. 

Therefore, we produce three sets of results: solution a) given in
Table~\ref{tab1} with estimated error of 0.1 \%; solution b) 
with error increased by five times to 0.5 \%; and c) in
Table~\ref{tab3} with realistic error of the GWU/SAID solutions estimated to
be 2 \%.

\begin{table*}[t!]
 \caption{List of results for resonance poles and branch points (bp) for the
   ideal case with errors of~0.1~\%.  
   Values in brakets denote estimated uncertainties.}
\label{tab1} %
\begin{tabular}{|c|c||c|c|c|}
\hline
   & & SM20 & AD19 & AD14  \tabularnewline
\hline
\hline
                 & real bp  &59.5(2) & 263.7(41.4) &435(314)  \tabularnewline
   $\chi^2_{df}$ & cmplx bp & 1.1(0.7) & 10.97(0.6) & 8.8(0.7) \tabularnewline
                 & real bp + 1-pole     & & 2.34(0.3) & 2.0(0.4)  \tabularnewline
\hline
resonance in & {\grey real b.p.}  & &  &   \tabularnewline
   3-body    & cmplx bp & 2260(22) - i 64(44) & 2352(1) - i 54(2)   & 2348(1) - i 48(2) \tabularnewline
sub-system   & {\grey real + bp 1-pole}     & &  &   \tabularnewline
\hline
  genuine    & {\grey real bp}  & &  &   \tabularnewline
2-body       & {\grey cmplx bp} & &  &  \tabularnewline
 resonance   & real bp + 1-pole     &  & \, 2362(0.7) - i 114(2) \, & \, 2362(0.6) - i 109(2) \,  \tabularnewline
\hline
\end{tabular}
\end{table*}


\begin{table*}[t!]
 \caption{Same as Table~\ref{tab1}, but for a realistic error of~2~\%. }
\label{tab3} %
\begin{tabular}{|c|c||c|c|c|}
\hline
   & & SM20 & AD19 & AD14  \tabularnewline
\hline
\hline
            & real bp  & 1.38(0.3)& 1.83(0.2) & 2.06(0.3)  \tabularnewline
    $\chi^2_{df}$ & cmplx bp & 0.97(0.16)& 0.99(0.16) & 0.98(0.16) \tabularnewline
            & 1-pole     &  & 1.03(0.15) & 1.08(0.5)  \tabularnewline
\hline
resonance in & {\grey real bp}  &  &   &    \tabularnewline
   3-body    & cmplx bp & 2265(76) - i 0 (7) & 2361(14) - i 59 (21) & 2354(12) - i 44(20) \tabularnewline
sub-system   & {\grey real bp + 1-pole}     &  &   &    \tabularnewline
\hline
  genuine    & {\grey real bp}  &  &   &    \tabularnewline
2-body       & {\grey cmplx bp} &  &   &   \tabularnewline
 resonance   & real bp + 1-pole     &  & \, 2361 (21) - i 63(20) \, & \, 2361(11) - i 60(13) \,  \tabularnewline
\hline
\end{tabular}
\end{table*}

What we immediately see from the tables is that the clearness of the
effect is the bigger the smaller the error bars are, and that is what we
expected.  A pole is certainly detected for AD14 and AD19 solutions, but it is
not clear, whether it is a real two-body resonance in two-body system
materialized as a genuine pole, or a two-body resonance  in the three-body
subsystem materialized as a complex branch-point. For the third SM20 solution
the possibility of the pole in a form of complex branch-point is preferred
only for ideal cases with unrealistically small error bars, but for realistic
error bars the situation is ambiguous.  For the smallest error bars in
Table~\ref{tab1} we see that all three solutions including SM20 require at
least a complex branch-point, but  for the realistic error in Table~\ref{tab3}
it is only clear that real branch-points are much less likely for all
three solutions ($\chi^2_{df}$ is the biggest, but not convincingly).  On the
other hand, for the ideal case given in Table~\ref{tab1} it is very likely
that these results could be interpreted only as a resonance in the two-body
system as $ \chi ^2 _{df}$  is notably higher for the other two possibilities
- real and complex branch-points. 
Our test with errors of 0.5 \% shows that already in this case 
equal probability for real pole and complex branch-point solution is reached. 
\\ \\
Therefore, we may conclude:

Both AD solutions require a pole in the system, however the distinction
between a pole in the two-body system or one in the three-body subsystem
depends on the reliability of the GWU/SAID solutions. The numerical
quantification also depends on the confidence limit of the GWU/SAID solutions.

However, if we add information from  sources other than just elastic $pn$
scattering, then the complex branch-point solution can be safely discarded. As
noted already above, the only possible 3-body branch point in the vicinity of
the found pole location is due to the $NN^*(1440)$ configuration. The Roper
resonance $N^*(1440)$ is much broader than suggested by the imaginary part of
the pole given in Table~\ref{tab3}. And since it is formed near threshold in
the isoscalar part of the $NN \to NN^*(1440)$ reaction preferentially by the $^3S_1$ $NN$
partial wave, a significant formation by the isoscalar $^3D_3-^3G_3$ partial
waves appears very unlikely. Finally, $d^*(2380)$, which may be
identified with the found pole, does not decay into $NN^*(1440)$ 
(BR < 14$\%$ at 90$\%$ C.L.) according to the recent measurement of the
isoscalar part of the $NN \to NN\pi$ reaction \cite{NNpi}.

In our tests with various different error assignments to the GWU/SAID
solutions the location of the pole position appears to be remarkable stable
against these 1error variations. The result of 2361(16) MeV is  compatible
with the traditional speed plot result of 2380(10) MeV \cite{npfull} within
uncertainties as well as with the result from the $np \to d\pi^0\pi^0$
reaction \cite{MB}, where a value of 2.37 GeV was observed
for the $d^*(2380)$ resonance energy.

The situation with respect to the resonance width is more delicate. Though the
width deduced with the L+P method still overlaps within uncertainties with
that deduced by the speed plot technique of 80(10) MeV, it is at notable
variance with the result of 70 MeV from $np \to d\pi^0\pi^0$. There are several
reasons for this discrepancy. First, as already noted in the discussion of the 
energy excitation function of the analyzing power around 90$^\circ$ ---  where
the resonance effect of $d^*(2380)$ is largest --- the measured resonance
structure at the high-energy side is narrower than accounted for by the AD
solutions. This failure causes a long high-energy tail of the resonance
structure seen in $Im(^3D_3)$, Fig.~\ref{FigGWU} top right. In consequence, the
resonance effect appears to more extended in the PW solutions than in the
data. Secondly, high-quality data beyond $\sqrt s$ = 2.44 GeV are rare in the
GWU/SAID data base and hence the uncertainties in the various PW solutions 
increase rapidly beyond this energy. {\it I.e.} the high-enery tail of the
$d^*(2380)$ resonance is not well fixed in the PW solutions causing a large
uncertainty in the separation of pole and background. This is particularily
true for the L+P method, where the resonance shape is kept unconstrained as
much as possible. Hence the true uncertainties for the imaginary part of the
pole appear to be even larger than given in Table~\ref{tab3}.








\section{Summary and Conclusions}

New data for the differential cross sections in the energy region of the
$d^*(12380)$ dibaryon resonance have been presented. They were extracted from
exclusive and kinematically complete measurements of quasifree $\vec{n}p$
scattering using the WASA detector setup at COSY and having a polarized
deuteron beam impinged on the hydrogen pellet target. The new cross section
data supplement the analyzing power data published already earlier
\cite{np,npfull}. 

The new cross section data are at obvious variance with the GW/SAID
partial-wave solutions SP07, SM16 and SM20, however, agree quantitatively with
the solutions AD14 and AD19. Whereas the first ones do not contain the
$d^*(2380)$ pole, the latter two do include this pole. The solution AD14 was
obtained 2014 by inclusion of the WASA analyzing power data into the SAID data
base. These data then produced 
the $d^*(2380)$ pole in the coupled $^3D_3-^3G-3$ coupled partial wave. It is
very gratifying and simultaneously demonstrates the predictive power of this
solution that it is able to provide a quantitative description of the new data
on the differential cross sections. The new solution AD19, which includes now
also the new cross section data in the SAID data base, deviates from the AD14
solution only marginally.

Since a looping in the Argand diagram is a necessary condition for a
resonance pole, but not yet a sufficient one, the three GWU/SAID solution AD14,
AD19 and SM20 were subjected to an interpretation within the
Laurent+Pietarinen method.
The conclusion there is that a pole at the position of the $d^*(2380)$
resonance is clearly confirmed. However understanding the effect as a
consequence of the 
$NN^*(1440)$ branch-point in the 3-body sub-channel can be excluded definitely
only by using additional information about the isoscalar part of the
$NN \to NN\pi$ reaction. Based on the elastic $pn$ scattering alone, a strict 
elimination of 
the branch-point interpretation would necessitate new precise high-quality
measurements, in particular at energies beyond $\sqrt s$ = 2.4 GeV, in order
to approach the precision given in Table~\ref{tab1}.

\section{Acknowledgments}

We acknowledge valuable discussions with C. Hanhart and C. Wilkin on this
issue. This work has been supported by BMBF, Forschungszentrum J\"ulich
(COSY-FFE) and the German Research Foundation
DFG (CL214/3-2 and 3-3). Three of us (H.~Cl., I.~S. and  A.~\v{S})
  appreciate the support
by the Munich Institute for Astro- and Particle Physics
(MIAPP) which is funded by the Deutsche Forschungsgemeinschaft (DFG, German
Research Foundation) under Germany's Excellence Strategy -- EXC-2094 -
390783311.
This work was
supported (W.B., I.S. and R.W.) in part by the U.~S.~Department of Energy,
Office of Science, Office of Nuclear Physics under Awards No. DE--SC0016583
and DE-SC0016582.

\end{document}